\documentclass[12pt, twocolumn]{iopart}
\usepackage{comment}
\usepackage{xcolor}
\usepackage{array}
\usepackage{arydshln}
\usepackage{amsmath}
\usepackage{subcaption}
\usepackage{gensymb}
\usepackage{cancel}

\usepackage{lipsum}
\usepackage{multirow}
\usepackage[hashEnumerators,smartEllipses]{markdown}

\usepackage[backend=biber,style=ieee,maxnames=4,urldate=comp,url=false]{biblatex}
\usepackage{float}
\usepackage{subcaption}
\usepackage{color,soul}
\usepackage{setspace}
\usepackage{import}
\usepackage{titlesec}
\usepackage[export]{adjustbox}
\usepackage{tocloft}
\usepackage{supertabular}

\usepackage{amsmath,physics,amssymb,slashed}
\usepackage{blkarray}
\usepackage{graphicx} 
\graphicspath{{./Images/}} 
\usepackage{caption} 

\usepackage{longtable} 

\usepackage[normalem]{ulem}
\useunder{\uline}{\ul}{}
\emergencystretch=\maxdimen
\usepackage{tikz}
\usetikzlibrary{calc,trees,positioning,arrows.meta,chains,shapes.geometric,  decorations.pathreplacing,decorations.pathmorphing,shapes,  matrix,shapes.symbols}
\tikzset{>={Latex[width=2mm, length=1.2mm]}}

\usepackage{float}

\usepackage{listings}
\usepackage{xcolor}

\usepackage{textcomp}

\definecolor{codegreen}{rgb}{0,0.6,0}
\definecolor{codegray}{rgb}{0.5,0.5,0.5}
\definecolor{codepurple}{rgb}{0.58,0,0.82}
\definecolor{backcolour}{rgb}{0.95,0.95,0.92}

\lstdefinestyle{mystyle}{
    backgroundcolor=\color{backcolour},   
    commentstyle=\color{codegreen},
    keywordstyle=\color{magenta},
    numberstyle=\tiny\color{codegray},
    stringstyle=\color{codepurple},
    basicstyle=\ttfamily\footnotesize,
    breakatwhitespace=false,         
    breaklines=true,                 
    captionpos=b,                    
    keepspaces=true,                 
    numbers=left,                    
    numbersep=5pt,                  
    showspaces=false,                
    showstringspaces=false,
    showtabs=false,                  
    tabsize=2,
}
\usepackage{pdfpages}

\usepackage{dirtytalk}
\usepackage[margin=.6in,includefoot,footskip=30pt]{geometry}
\usepackage{makecell}
\usepackage{graphicx,booktabs}
\usepackage{multicol}
\AtEveryBibitem{\clearfield{issn}}
\AtEveryBibitem{\clearfield{eprint}}
\AtEveryBibitem{\clearfield{isbn}}

\begin{document}

\twocolumn[
  \begin{@twocolumnfalse}
    \title{MANTA: A Negative-Triangularity NASEM-Compliant Fusion Pilot Plant}
\author{%
The MANTA Collaboration,
G.~Rutherford$^{1}$,
H.~S.~Wilson$^{2}$,
A.~Saltzman$^{1}$,
D.~Arnold$^{2}$,
J.~L.~Ball$^{1}$,
S.~Benjamin$^{1}$,
R.~Bielajew$^{1}$,
N.~de~Boucaud$^{3}$,
M.~Calvo-Carrera$^{1}$,
R.~Chandra$^{2}$,
H.~Choudhury$^{2}$,
C.~Cummings$^{1}$
L.~Corsaro$^{1}$,
N.~DaSilva$^{2}$,
R.~Diab$^{1}$,
A.~R.~Devitre$^{1}$,
S.~Ferry$^{1}$,
S.~J.~Frank$^{1}$,
C.~J.~Hansen$^{2}$,
J.~Jerkins$^{1}$,
J.~D.~Johnson$^{1}$,
P.~Lunia$^{2}$,
J.~van~de~Lindt$^{1}$,
S.~Mackie$^{1}$,
A.~D.~Maris$^{1}$,
N.~R.~Mandell$^{1}$,
M.~A.~Miller$^{1}$,
T.~Mouratidis$^{1}$,
A.~O.~Nelson$^{2}$,
M.~Pharr$^{2}$,
E.~E.~Peterson$^{1}$,
P.~Rodriguez-Fernandez$^{1}$,
S.~Segantin$^{1}$,
M.~Tobin$^{2}$,
A.~Velberg$^{1}$,
A.~M.~Wang$^{1}$,
M.~Wigram$^{1}$,
J.~Witham$^{1}$,
C.~Paz-Soldan$^{2}$,
and
D.~G.~Whyte$^{1}$
}

\address{$^{1}$~Plasma Science and Fusion Center, Massachusetts Institute of Technology, Cambridge, MA 02139, USA}
\address{$^{2}$~Department of Applied Physics and Applied Mathematics, Columbia University, New York, NY 10027, USA}
\address{$^{3}$~General Atomics, San Diego, CA 92121, USA}

\maketitle
\begin{abstract}
The MANTA (Modular Adjustable Negative Triangularity ARC-class) design study investigated how negative-triangularity (NT) may be leveraged in a compact, fusion pilot plant (FPP) to take a ``power-handling first" approach. The result is a pulsed, radiative, ELM-free tokamak that satisfies and exceeds the FPP requirements described in the 2021 National Academies of Sciences, Engineering, and Medicine report ``Bringing Fusion to the U.S. Grid"\supercite{NASEM_report}. A self-consistent integrated modeling workflow predicts a fusion power of 450 MW and a plasma gain of 11.5 with only 23.5 MW of power to the scrape-off layer (SOL). This low $P_\text{SOL}$ together with impurity seeding and high density at the separatrix results in a peak heat flux of just 2.8 MW/m$^{2}$. MANTA's high aspect ratio provides space for a large central solenoid (CS), resulting in ${\sim}$15 minute inductive pulses. In spite of the high B fields on the CS and the other REBCO-based magnets, the electromagnetic stresses remain below structural and critical current density limits. Iterative optimization of neutron shielding and tritium breeding blanket yield tritium self-sufficiency with a breeding ratio of 1.15, a blanket power multiplication factor of 1.11, toroidal field coil lifetimes of $3100 \pm 400$ MW-yr, and poloidal field coil lifetimes of at least $890 \pm 40$ MW-yr. Following balance of plant modeling, MANTA is projected to generate 90 MW of net electricity at an electricity gain factor of ${\sim}2.4$. 
Systems-level economic analysis estimates an overnight cost of US\$3.4 billion, meeting the NASEM FPP requirement that this first-of-a-kind be less than US\$5 billion. The toroidal field coil cost and replacement time are the most critical upfront and lifetime cost drivers, respectively.
\end{abstract}
  \end{@twocolumnfalse}
]


\section{Introduction and Overview}
\label{sec:intro}

For fusion energy to contribute towards achieving decarbonization targets, a fusion pilot plant (FPP) must begin operation in the 2030s. Scoping an FPP must therefore begin now,~\supercite{NASEM_report} and its design must provide cost and operational certainty for fusion energy commercialization. A significant challenge in scaling the tokamak concept, and more generally magnetic fusion, to an FPP is maintaining a high performance core plasma with peak heat fluxes to the plasma facing components (PFCs) within technological limits for heat removal. And this must be achieved simultaneously with the high average power desired for commercial fusion devices. Exceeding this heat removal limit would result in component failure and a maintenance period for replacement. Furthermore, erosion of PFCs occurs even in modern devices\supercite{Diez_2021, Huber_2021}, where discharges durations and steady-state heat flux fluxes are, respectively, ${\sim}2$ and ${\sim}1$ orders of magnitude below that expected for reactors\supercite{Litaudon_2024,kuang_divertor_2020, menard_fusion_2022}. FPPs and commercial fusion devices will therefore require a dissipative divertor, where a significant amount of power is radiated between the separatrix and the divertor targets, and overall a reliable and robust strategy to dissipate fusion plasma power to the PFCs. 

Typical H-mode operation results in an additional transient heat flux in the form of edge-localized modes (ELMs)\supercite{kuang_divertor_2020,menard_fusion_2022}. These bursts of particles and energy through the SOL pose a significant risk to PFCs, and their severity increases with the stored energy of the plasma\supercite{HILL1997182}. Proposed devices such as EU-DEMO will be unable to operate without ELM control\supercite{DEMO_elm_limitations}. While ELM-free regimes do exist, the ability to concurrently achieve high performance, ELM mitigation, and a dissipative divertor is often restricted to subsets parameter space, which may not be compatible with the requirements of a power plant\supercite{Paz-Soldan_2021, VIEZZER2023101308}.

\begin{figure}
\centering
\includegraphics[width=\columnwidth]{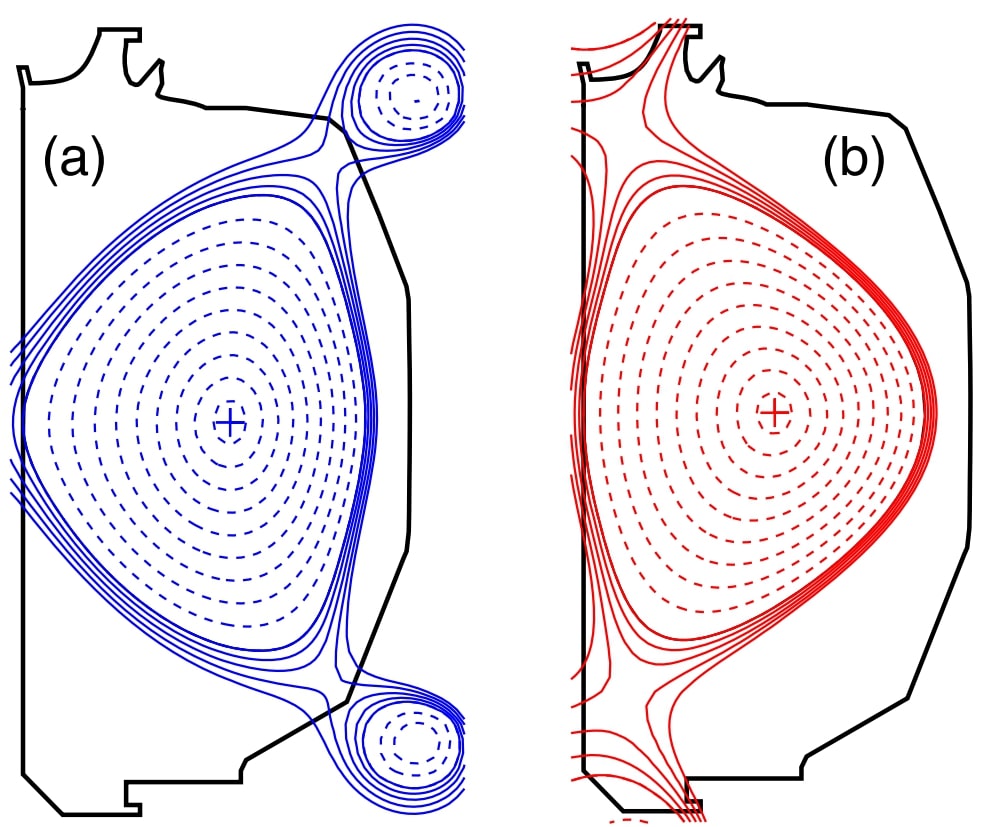}
\caption{Comparison of the plasma cross sections for negative (a) and positive (b) triangularities on the DIII-D tokamak. Figure reproduced from \cite{austin2019achievement} with the permission of the American Physical Society.}
\label{fig:pt_vs_nt}
\end{figure}

A potential solution is the use of negative triangularity ($\delta < 0$), where the plasma's D-shaped cross section is inverted relative to the usual positive triangularity. The triangularity $\delta$ is taken to be the average up the upper and lower triangularities $\delta_{u,l} = (R_\text{geo} - R_{u,l})/a$, where $R_\text{geo}$ is the geometrical major radius, $R_\text{u,l}$ is the major radius of the highest/lowest point on the last closed flux surface (LCFS), and $a$ is the minor radius\supercite{Luce_2013}. A comparison between a positive and negative triangularity plasma with otherwise identical shaping parameters is given in Fig \ref{fig:pt_vs_nt}. With sufficiently negative triangularity (NT), the 2nd stability region for infinite-n ballooning modes is inaccessible\supercite{Nelson_2022, Nelson_2023}. This prevents the plasma from entering H-mode and developing the steep edge gradients that produce ELMs. There is also then no requirement of a minimum SOL power as there is in H-mode. This permits a higher fraction of the total power to be radiated from the core, reducing the power incident on the divertor targets. Additional power-handling benefits are detailed in Section \ref{sec:divertor}. It is important to note that even without H-mode, the use of NT does not preclude high fusion performance, as will be discussed in Section~\ref{sec:core}.

Application of NT to power-plant scale tokamaks has been explored in several previous works\supercite{Kikuchi_2014, medvedev2016single,Kikuchi_2019, Medvedev_2015}. This paper builds on previous studies but differs in two major ways. First, all magnets are assumed to be constructed from Rare-Earth Barium Copper Oxide (REBCO) high-temperature superconductors (HTS), opening portions of parameter space not previously investigated for NT designs. Second, an integrated reactor design is presented, in which core transport, power-handling, magnet systems, neutronics, and economic viability were simultaneously optimized to inform the overall design. 

\begin{figure}
\centering
\includegraphics[width=\columnwidth]{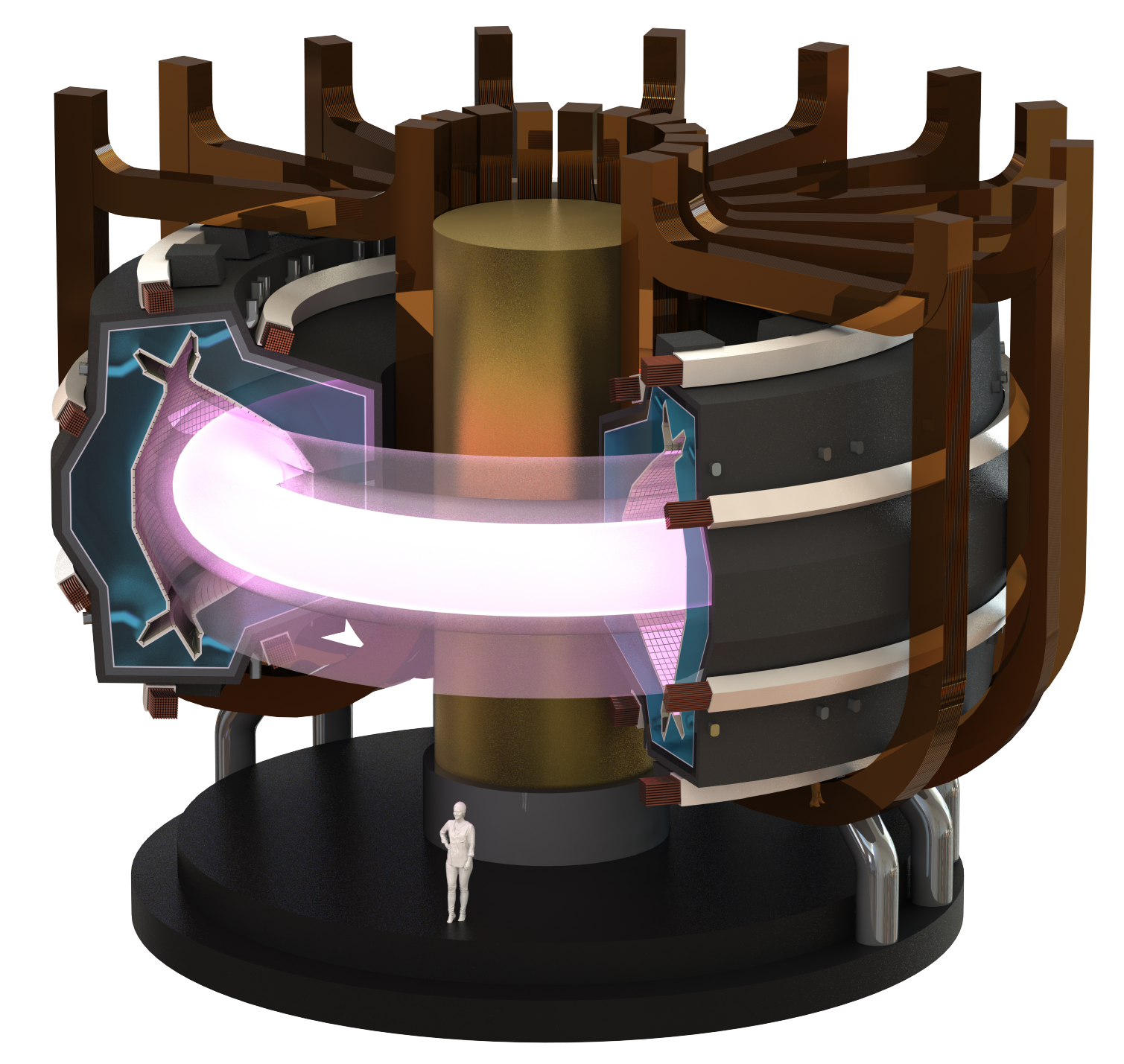}

\caption{MANTA is a compact, negative triangularity fusion pilot plant. An average sized human adult is present for scale.}
\label{fig:MANTA_CAD}
\end{figure}

The result of this study is MANTA (Modular Adjustable Negative-Triangularity ARC-class): a pulsed, radiative, ELM-free, negative triangularity ARC-class\supercite{ARC_2015} tokamak FPP. MANTA's design is shown in Fig \ref{fig:MANTA_CAD} and was developed with a focus on maximizing self-consistency. MANTA advances the three scientific and technological readiness drivers outlined in the Fusion Energy Sciences Advisory Committee's ``Powering the Future Fusion \& Plasmas" report\supercite{FESAC_report}: sustaining a burning plasma, engineering for extreme conditions, and harnessing fusion energy. MANTA also surpasses the criteria to demonstrate the path to commercial viability of nuclear fusion energy as laid out in the National Academies of Sciences, Engineering, and Medicine (NASEM) ``Bringing Fusion to the U.S. Grid" report\supercite{NASEM_report}, namely:
\begin{enumerate}
    \item Electricity gain factor: $Q_\text{e} > 1$
    \item Continuous net electricity $\geq$  50~MWe for at least 3 hours
    \item Tritium Breeding Ratio (TBR) $\gtrsim$ 0.9
    \item Overnight cost $<$ US$\$5$B
    \item Operation through several environmental cycles 
\end{enumerate}

It is worth emphasizing that MANTA is not a commercial power plant, but rather a pilot plant. Maximizing absolute performance was therefore not the goal of this study. Instead, a focus was placed on maintainability and flexibility at the conditions relevant to a power plant. Further optimizations may be possible, but MANTA is well-suited to its role as a pilot plant and both fulfills the NASEM requirements and advances the FESAC goals.

Namely, MANTA's modularity and rapid maintenance permits faster advancement of fusion science and technology. This is achieved through a liquid immersion FLiBe tank that permits modifications to the vacuum vessel (VV) without requiring changes to the blanket system and demountable toroidal field (TF) coils that give ready access to the FLiBe tank and VV. The TF coils may be ramped relatively quickly due to an oversized cryosystem, and replacement of the FLiBe tank and VV as a single assembly reduces radiation hazards and speeds maintenance.

Additionally, while only a single operating point was investigated with integrated modeling, 0D scoping predicts adjustable fusion power with near constant $P_\text{SOL}$ through control of the density. This allows for the testing of physics and technology over a range of conditions without worsening the power-handling challenge. 

\begin{table}[h!]
\begin{center}
\caption{MANTA Key Design Parameters}
\label{tab:MANTA_params}
\resizebox{\columnwidth}{!}{
\begin{tabular}{lll}
 \hline
 Parameter & Symbol & Value\\
 \hline
 \hline
Fusion power & $P_\text{fus}$ & 450 MW\\
Total thermal power & $P_{\rm th}$ & 530 MW\\
Net electric power & $P_{\rm e,net}$ & 90 MWe\\
ICRF coupled power & $P_{\rm ICRH}$ & 40 MW\\
Scrape-off layer power & $P_\text{SOL}$ & 23.5 MW\\
Plasma gain & $Q$ & 11.5\\
Electricity gain & $Q_E$ & 2.4\\
Major radius & $R_0$ & 4.55 m\\
Plasma minor radius & a & 1.2 m\\
Plasma elongation & $\kappa$ & 1.4\\
Plasma triangularity & $\delta$ & -0.5\\
Plasma volume & $V_p$ & 155 $m^{3}$\\
Plasma surface area & $A_p$ & 258 $m^{2}$\\
Toroidal magnetic field & $B_0$ & 11 T\\
Plasma current & $I_P$ & 10 MA\\
Bootstrap fraction & $f_{\rm BS}$ & 18\%\\
Tritium breeding ratio & TBR & 1.15\\
Avg. ion temperature & 〈$T_i$〉& 7.3 keV\\
Avg. electron temperature & 〈$T_e$〉&  7.1 keV\\
Avg. density & 〈n〉 & 1.95$\cdot 10^{20}$ m$^{-3}$\\
On-axis ion temperature & $T_{i,0}$ & 19 keV\\
On-axis e$^{-}$ temperature & $T_{e,0}$ & 18.8 keV\\
On-axis e$^{-}$ density & $n_0$ & 2.76$\cdot 10^{20}$ m$^{-3}$\\
Greenwald fraction & $f_{\rm Gr}$ & 0.88\\
Pulse length & $\tau_{\rm pulse}$ & 15 min\\
Inter-pulse length & $\tau_{\rm inter}$ & 2 min\\
Normalized beta & $\beta_N$ & 1.35\\
Safety factor at $\Psi_{N}$ = 0.95 & $q_{95}$ & 2.3\\
Minimum safety factor & $q_{\rm min}$ & 0.905\\
Energy confinement time & $\tau_{E}$ & 0.94 s\\
H$_{89}$ confinement factor & H$_{89}$ & 1.44\\
Loop voltage  & V$_\text{loop}$ & 0.206 V\\
 \hline
\end{tabular}
}
\end{center}
\end{table}

A list of MANTA's key parameters at its design point is given in Table \ref{tab:MANTA_params}. The remainder of the paper consists of the following sections:
Sec.~\ref{sec:core} scopes MANTA's fusion core solution via 0-D power balance before refining the operating point with an integrated modeling workflow;
Sec.~\ref{sec:divertor} develops a power-handling solution capable of withstanding the exhausted and radiated power determined by the core modeling;
Sec.~\ref{sec:magnets} details the design, mechanical analysis, and maintenance of the toroidal field (TF) coils, central solenoid (CS), and poloidal field (PF) coils;
Sec.~\ref{sec:neutronics} uses neutronics simulations to calculate magnet lifetimes and analyze the tritium fuel cycle;
Sec.~\ref{sec:balance_of_plant} discusses MANTA's balance of plant and determines the steady state net electrical power;
Sec.~\ref{sec:econ} provides an economic analysis of MANTA to investigate its financing; and 
Sec.~\ref{sec:conclusion} gives concluding remarks.

\section{Plasma Core}
\label{sec:core}

In addition to its advantages in power-handling, negative triangularity also improves core performance with some NT plasmas on DIII-D and TCV achieving H-mode-level confinement (H$_{98}$ = 1).\supercite{austin2019achievement, Coda_2022, Marinoni_2021, Marinoni_2019, pazsoldan2023} This is made possible by NT's weaker power degradation of the energy confinement time and a reduction in the electron heat transport.\supercite{Marinoni_2021, Camenen_2007, Marinoni_2019, Fontana_2020} Importantly, high performance dimensionless parameters of $\beta_N > 3$, $f_{Gr} > 1$, and $q_{95} < 3$ have been achieved simultaneously in a diverted configuration with this high confinement of $H_{98} > 1$\supercite{pazsoldan2023}, where $f_{Gr}$ is the Greenwald fraction, $\beta_N$ is the normalized $\beta$, and $q_{95}$ is the safety factor at $\psi_N = 0.95$. These promising results warrant the investigation of NT FPP designs, like MANTA, despite NT being far less explored than PT. 

\subsection{0-D Scoping of the Core Solution}
Broad scoping of the space of core solutions capable of meeting the NASEM requirements was completed through the use of POPCONs (Plasma OPerational CONtours), 0-D models of tokamak core performance that solve the global power-balance equation~\supercite{Houlberg_1982}. MANTA's POPCONs were generated with the open source code CFSpopcon~\supercite{cfspopcon}. Key POPCON outputs include fusion power $P_\mathrm{fus}$, required auxiliary heating power $P_{\rm aux}$, radiated power $P_{\rm rad}$, and plasma gain $Q_p$, all of which were calculated over a range of volume-averaged $T_e$ and $n_e$ values. The temperature and density profiles were taken as user-inputs. Krypton was chosen as the core radiator to achieve the necessary radiated power, primarily in the outer parts of the core plasma, to maintain low $P_\mathrm{SOL}$. The krypton density profile was assumed to be a scaled version of the electron density profile. It was additionally assumed that the radiative power fractions viable in PT L-mode\cite{ARCH_2022} were also acceptable for ELM-free NT. Uncertain parameters, such as the confinement factor or profile shaping parameters were varied over conservative ranges from the literature. Historical PT L-mode data~\supercite{angioni_relationship_2005} was used as reference for density peaking, while the recent DIII-D NT campaign \supercite{pazsoldan2023} informed the choice of H98y2 energy confinement scalings~\supercite{iter_physics_expert_group_on_confinement_and_transport_chapter_1999}.

Additional free parameters included the triangularity $\delta$, the elongation $\kappa$, the minor radius $a$, the major radius $R_0$, the toroidal magnetic field $B_T$, and the plasma current $I_p$. To ensure the second stability region for infinite-n ballooning modes and hence H-mode was inaccessible regardless of the choice of other parameters, a highly negative triangularity of $\delta = -0.5$ was selected. $B_T$ on axis was assumed to be 11 T as this is broadly consistent with the magnetic field in other high-field tokamak designs, such as SPARC\supercite{SPARCbasis}, which have completed extensive magnet analyses. $I_p$ was chosen to be driven inductively to avoid the cost and complexity of a non-inductive current drive system. This leads to a preference for larger aspect ratios as the wider inner bore fits a larger central solenoid capable of a greater flux swing. An increased magnetic flux extends the duration of flattop operation, aiding MANTA's ability to meet the fusion power duration requirement.

To satisfy the NASEM requirement of 50~MWe net electric power, the minimum $P_\mathrm{fus}$ was initially estimated to be $200$~MW (though this would later prove to be too low), with higher $P_\mathrm{fus}$ desirable within the constraints of component lifetimes and economics. $P_\mathrm{fus}$ increases with larger $I_p$ due to the longer confinement time\supercite{iter_physics_expert_group_on_confinement_and_transport_chapter_1999} and with $R_0$, $a$, and $\kappa$ due to both increases in the confinement time and also the larger plasma volume: $V_\text{p}\approx 2\pi^2 R_0 a^2 \kappa$. $I_p$ is constrained by the increased risk of kink instability at large currents. To account for this, $I_p$ was removed as a free parameter and replaced by a target kink safety-factor $q_* \approx 2.5$, well above the marginal kink safety factor $q_\star=2$ reported in \cite{menard_aspect_2004}. $I_p$ was then calculated from $q_*$ as per \cite{menard_aspect_2004}:
\begin{equation}\label{current_scaling}
I_p  = \frac{2\pi a^2 B_0}{\mu_0 R_0 q_*}\Bigl(\frac{1+\kappa^2}{2}\Bigr)10^{-6} \text{ [MA]}.
\end{equation}

$\kappa$ was constrained by the increased risk of vertical displacements at high $\kappa$\supercite{freidberg_tokamak_2015}, an effect exacerbated by NT\supercite{song_impact_2021}. To balance plasma volume and stability, $\kappa = 1.4$ was selected based on results from the AVSTAB vertical stability model~\supercite{song_impact_2021,Nelson_vertControl}. The use of passive stability plates may allow for access to higher elongations\supercite{guizzo2024assessment}, but this was not explored for MANTA's design and could be an area of future work.

$R_0$ and $a$ were constrained by the expense of the toroidal field magnets (which scales with the plasma surface area $\approx 2\pi^2R_0 a(1+\kappa)$) and by component lifetimes. Using the Monte Carlo code OpenMC\supercite{Romano2015}, neutronics simulations of some POPCON solutions were completed. The magnetic equilibria of these cases were computed by the Grad-Shafranov solver CHEASE~\supercite{LutjensCHEASE} and were to assumed to be up-down symmetric double nulls. Such an equilibrium both aids in power-handling and simplifies MANTA's design. From these simulations and economic calculations, $R_0=4.55$ and $a = 1.2$ were selected to balance performance, component lifetime, and device cost. $I_p$ was then set at $10$~{ MA}, resulting in $q_\star=2.59$. It is important to note that the avoidance of tearing modes was not factored into parameter selection and could be an area of future work. 

\begin{figure}[h]
\centering
\includegraphics[width=\columnwidth]{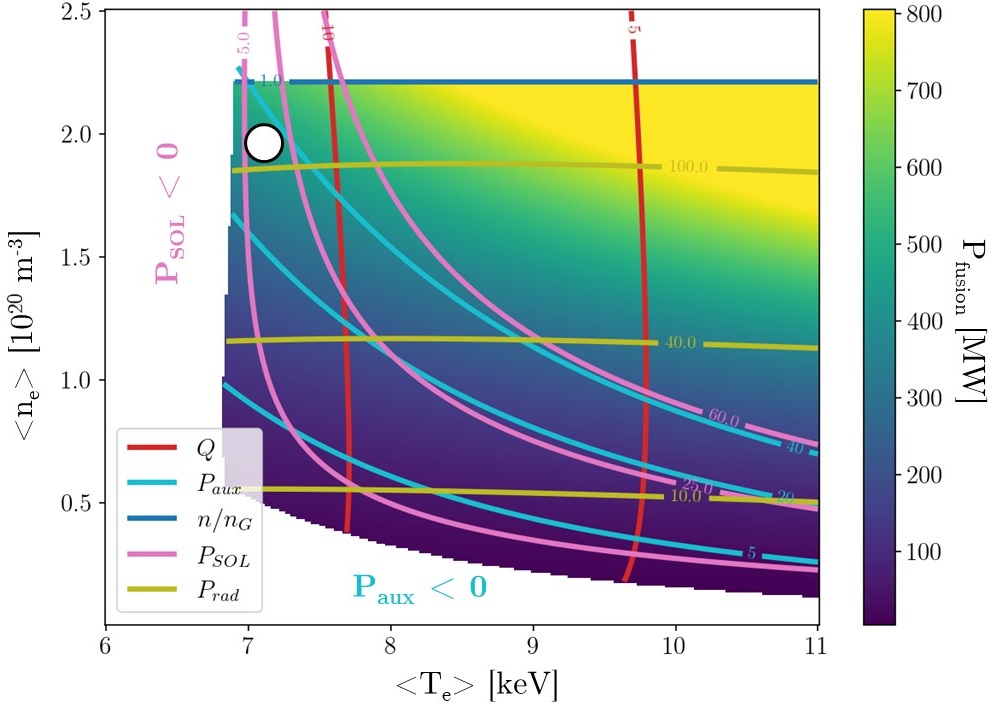}
\caption{POPCON analysis of the parameter space around MANTA's operating point. The chosen operating point is marked with a white circle, and the possible operating space is the colored region. At higher densities, the Greenwald density limit is exceeded. Lower temperatures and densities result in non-physical solutions. The profiles and confinement used in this POPCON matched those from the STEP modeling detailed in Sec. \ref{sec:STEP}.}
\label{fig:POPCON}
\end{figure}

The POPCON for this scenario is given in Fig.~\ref{fig:POPCON}, where MANTA's operating point is marked by the white circle, and the density and temperature profiles are those calculated by the transport code TGYRO~\supercite{CandyTGYRO}, detailed below. Notably, this POPCON features near vertical contours of $P_\mathrm{SOL}$ near the chosen operating point. A range of $P_\mathrm{fus}$ values can therefore be achieved with near constant $P_\mathrm{SOL}$ through controlling the density. This adjustability permits the potential study of multiple operating points without largely affecting the divertor power handling solution. And the choice of an inductive plasma means these changes in density affect the pulse length, rather than the ability to drive the required current, as would be the case in a non-inductive plasma. 

An ion cyclotron range of frequencies (ICRF) minority heating \supercite{Stix_minority} system was selected to supply the required 40 MW of auxiliary power for this operating point due the efficient bulk core heating such a system provides. Mirroring SPARC~\supercite{SPARC_RF}, $\rm ^3He$ was chosen as the minority species, which takes advantage of the overlap between the fundamental and second harmonic resonances of $\rm ^3He$ and tritium, respectively. A minority fraction of \mbox{$f_{\mathrm{He3}} = n_{\mathrm{He3}}/n_{\mathrm{e}} = 2.5$\%} was taken to maintain good wave polarization with dilution away from the optimal 50/50 D-T ratio. The launched parallel refractive index $N_{\parallel}$ was chosen to be similar to SPARC \supercite{SPARC_RF}, resulting in a toroidal refractive index $N_{\phi} = 30$. On-axis damping corresponded to a frequency of 110 MHz, readily achievable by existing high power tetrodes \supercite{Cmod_engineering}. While a detailed antenna design is outside the scope of this study, it should be noted that NT places the broad side of the plasma on the outboard wall, giving a large, relatively flat area suitable to a variety of antenna configurations. 

\subsection{Plasma Core Integrated Workflow} \label{sec:STEP}

\begin{figure}[h!]
\centering
\includegraphics[width=\columnwidth, trim={0 0 10cm 0}, clip]{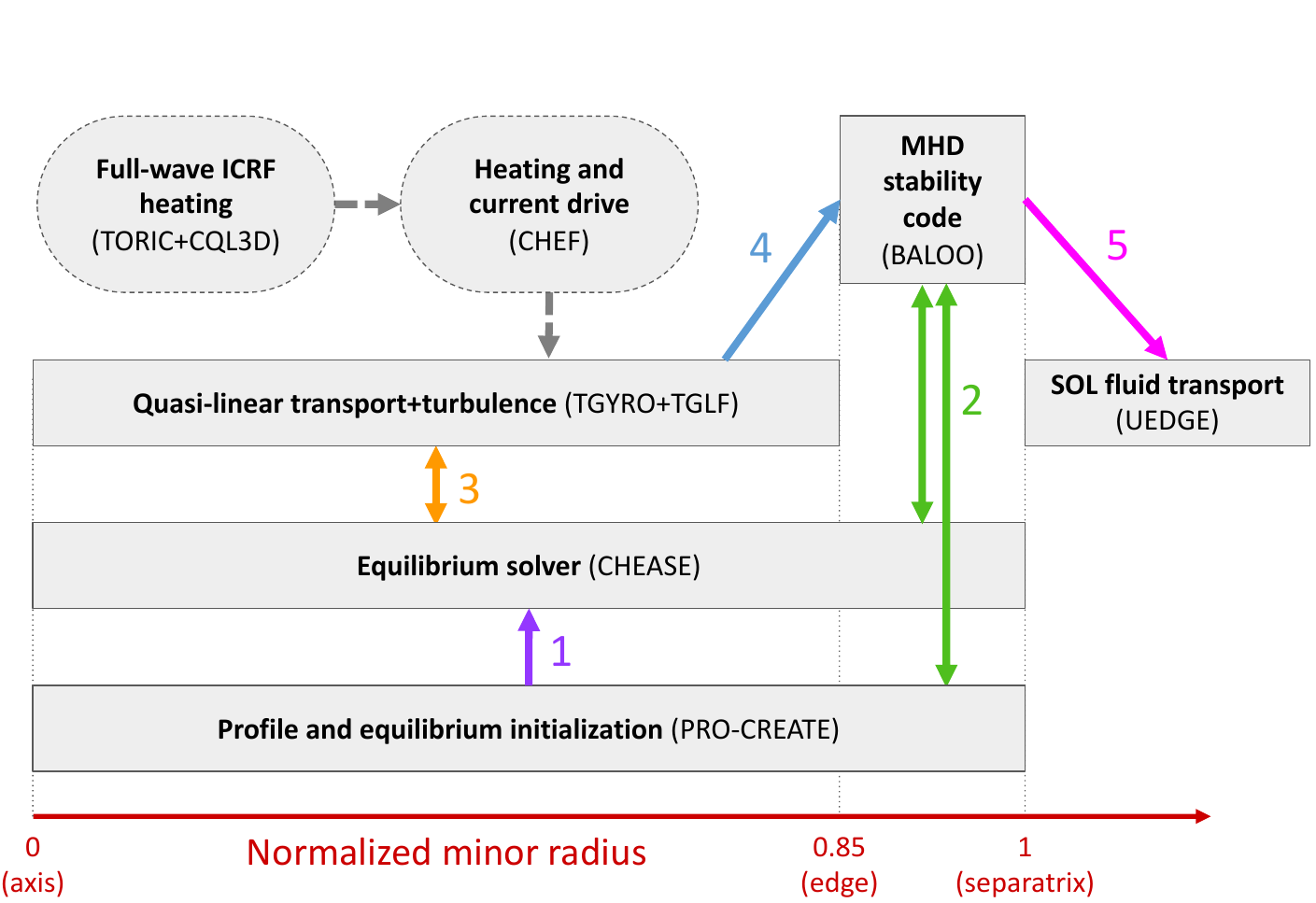}
\caption{Integrated workflow schematic showing the codes used to simulate the various regions of the plasma and the iteration between these codes.} 
\label{fig:integrated_workflow}
\end{figure}

Following selection of MANTA's 0-D parameters from the POPCON scoping, an integrated workflow was employed to obtain a more realistic and self-consistent core solution. This workflow, illustrated in Fig.~\ref{fig:integrated_workflow}, linked multiple high-fidelity codes to map plasma parameters from the core to the divertor targets. This was made possible in part by the OMFIT STEP module~\supercite{MeneghiniSTEP}, which allows self-consistent iteration between equilibrium, heating, transport, and stability codes in the OMFIT framework~\supercite{MeneghiniOMFIT}. STEP accomplishes this via the OMAS  (Ordered Multidimensional Array Structure) data structure to automatically transfer the outputs of one OMFIT module as inputs of the next. 

The codes used in this workflow and their functions are as follows: CHEASE\supercite{LutjensCHEASE} is a fixed-boundary Grad-Shafranov solver that produced MHD magnetic equilibria; PRO-create\supercite{procreate} generated plasma profiles from 0D parameters; BALOO\supercite{MillerBALOO} calculated the ballooning stability of the edge pressure gradient; TORIC\supercite{TORIC} is a full-wave solver that determined the electromagnetic fields resulting from the ICRF system; CQL3D\supercite{cql3d} is a Fokker-Plank code that evolved the distribution functions of the chosen species, in this case due to the RF; CHEF\supercite{LyonsCHEF} runs multiple heating and current drive codes to predict steady-state power deposition and current density profiles; TGYRO\supercite{CandyTGYRO} is a transport code that evolved plasma profiles such that collision and transport losses balance the input power; The neoclassical transport code NEO \supercite{BelliNEO} calculated these collisional losses, and the quasilinear turbulent transport code TGLF \supercite{StaeblerTGLF} calculated the transport losses; and UEDGE is a 2D edge transport code that extended plasma profiles to the divertor targets (see Section \ref{sec:divertor}). Throughout this workflow, density profiles were held fixed with a peaking value given by the Angioni scaling\supercite{Angioniscaling}. This was done primarily to simplify flux matching performed by TGLF by excluding particle flux, and it is of note that particle sources in a reactor are expected to be localized outside of $\rho = 0.95$ and TGLF has shown significant variability in density peaking prediction\supercite{rodriguezfernandez2022}. Allowing density to also evolve may be investigated in future work.

The difficulty in maintaining self-consistency throughout this workflow is that all of the codes are functions of the plasma profiles, but these profiles are the result of running all of the codes. For this reason, the workflow was completed twice. The first iteration began with passing CHEASE's magnetic equilibrium (generated for the POPCON scans) to STEP. Density and temperature profiles were then initialized in PRO-create\supercite{procreate} based on the core temperature and density predicted by the POPCON analysis under the constraint of edge conditions (from BALOO~\supercite {MillerBALOO}) feasible for a ballooning stable NT edge. Using these density and temperature profiles, an updated equilibrium was generated. The new CHEASE equilibrium, PRO-create plasma profiles, and CHEF heating sources were then fed into TGYRO, which evolved the electron and ion temperature profiles. TGYRO and CHEASE were iterated between until convergence was reached. An in-depth description of the TGYRO transport simulations and PRO-create profile generation is given in subsection~\ref{sec:TGYRO}. 

The second iteration passed the TGYRO temperature profiles and the PRO-create density profiles to TORIC/CQL3D (further described in Section \ref{sec:TORIC_CQL}) to generate more accurate auxiliary heating profiles for the bulk ions and electrons. These heating profiles were copied into CHEF and updated temperature profiles were calculated by TGYRO. Again TGYRO was iterated with CHEASE until convergence. The profiles resulting from this workflow were then passed to UEDGE to extend the solution from the separatrix to the divertor targets (see Sec.~\ref{sec:divertor}). Future work could entail further iterations of this workflow to improve self-consistency. 

\subsubsection{RF Power Deposition Profiles from TORIC/CQL3D} 
\label{sec:TORIC_CQL}
\hfill

\begin{figure}[h]
\centering
\includegraphics[width=\columnwidth]{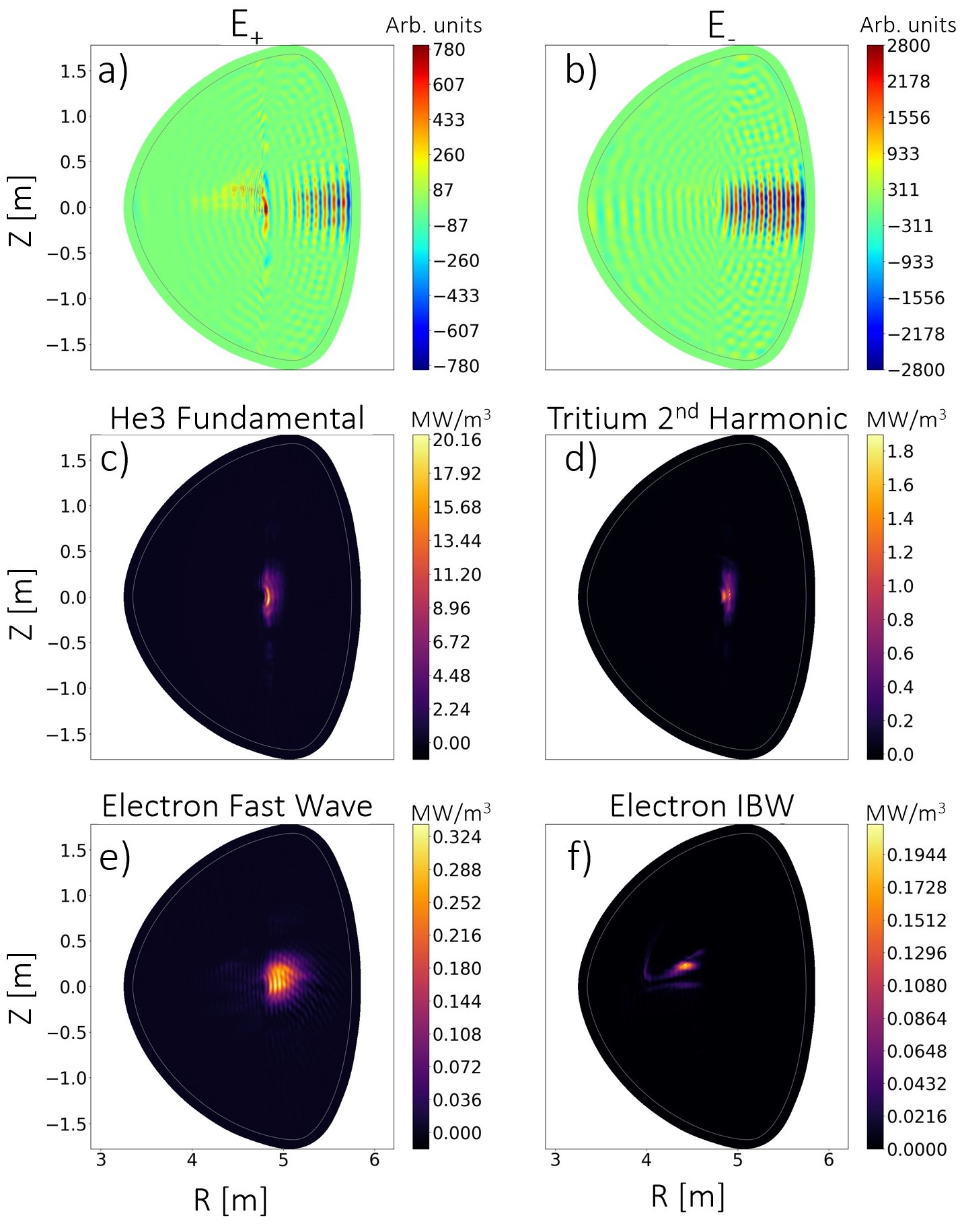}
\caption{The electric field and RF power deposition calculated by TORIC/CQL3D. There is little direct absorption of the RF power by the electrons. (a) and (b) show the two circularly polarized electric field components ($E_{||}$ is small and is not shown). (c) and (d) show the RF power deposition for the $\rm ^3He$ minority and T, respectively. (e) and (f) show the RF power deposition for the the electrons due the fast wave and the ion Bernstein wave, respectively.}
\label{fig:toric_plot}
\end{figure}

The full-wave code TORIC \supercite{TORIC} solved Maxwell's equations to calculate the electric fields generated by the ICRF system and obtain the RF power deposition profiles for D, T, $\rm ^3He$, Kr, and electrons. The resulting electric field and power deposition cross sections are shown in Fig.~\ref{fig:toric_plot}. (a) and (b) show the two circularly polarized electric field component magnitudes.
The choice of $f_{\mathrm{He3}} = 2.5$\% resulted in a large value of $E_+$ at the fundamental $\rm ^3He$ resonance, producing strong local heating of up to 20 MW/m$^3$, as shown in (c). Being collocated at this region of large $E_+$, the next strongest heating is 2nd harmonic tritium, shown in (d). Electron heating due to the fast wave and the ion Bernstein wave (IBW), shown in (e) and (f), are relatively small. Heating of D and Kr were found to be negligible and are not shown.

The Fokker-Planck code CQL3D\supercite{cql3d} was then coupled\supercite{sam_frank_aps_talk_2023} to TORIC to determine how the energy deposited on the minority ions was partitioned to the bulk ions and electrons via Coulomb collisions. The two codes were iterated 100 times with good convergence. The resulting 1D power deposition profiles calculated by TORIC/CQL3D (shown in Figure \ref{fig:TGYROprofiles} as $Q_{\text{e, aux}}$ and $Q_{\text{i, aux}}$) were then passed to CHEF for easy integration into STEP for the second iteration of TGYRO.  

\subsubsection{Transport Simulation with TGYRO in STEP}\label{sec:TGYRO}\hspace{40ex}
Transport modeling was completed with TGYRO, which modified the temperature profiles such that the collision and transport losses, calculated respectively by the neoclassical transport code NEO \supercite{BelliNEO} and the quasilinear turbulent transport code TGLF \supercite{StaeblerTGLF}, balance the input power. TGLF was run with saturation rule SAT-2\supercite{Sat2_paper1, Sat2_paper2}, as this produces profiles that are in good agreement with DIII-D NT experiments \supercite{McClenaghan2024} and has been used in similar conceptual design studies\supercite{ARCH_2022}. Species included in the simulation were electrons, D, T, and Kr. Notably, $2.5\%$ $^3$He and ${\sim}2\%$ $^4$He were not included and could be an area of future research. Bremsstrahlung and synchotron radiation are included. Temperature profiles with various core and edge values were converged until a solution was found with acceptable scrape-off-layer power $P_\text{SOL}$, radiative fraction $f_\text{rad}$, fusion power $P_\text{fus}$,  and Greenwald faction $f_\text{Gr}$. 

Free variables for STEP profile optimization were electron density at the boundary of TGYRO evolution ($\rho = 0.85$) $n_{85}$, Kr concentration $f_{\text{Kr}}$, and temperature at $\rho = 0.85$ $T_{85}$. D and T density profiles were generated by PRO-create assuming a 50/50 mix and enforcing quasineutrality given the electron and Kr density profiles. The Kr density profile was assumed to be the same shape as the electron density profile, justified by the L-mode-like particle transport expected in NT. Edge temperature was assumed to be the same for all species. Edge values were defined to be at $\rho_\text{tor} = 0.85$. 

\begin{figure}[h]
\centering
\includegraphics[width=\columnwidth]{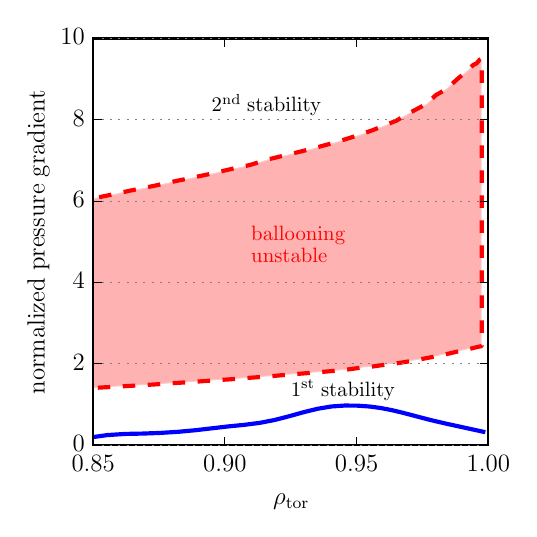}
\caption{The equilibrium pressure gradient (blue curve, normalized) is prevented from growing steep due to the region of ballooning instability (pink region), calculated by the code BALOO \supercite{MillerBALOO}.} 
\label{fig:BALOO_plot}
\end{figure}

Initial scoping with UEDGE (described in Section \ref{sec:divertor}) found that a separatrix density of $n_{\text{sep}} = 0.9\times 10^{20}$ m$^{-3}$ allowed for sufficiently low heat flux on the divertor targets, giving a lower bound for $n_{85}$. The on-axis density $n_0$ was set such that density peaking followed the Angioni scaling. The upper bound for $n_{85}$ was set by enforcing that the pressure gradient around $\rho_{\rm tor} = 0.85$ remained well below the ballooning stability limit (Fig. \ref{fig:BALOO_plot}) and that the volume average density did not exceed the Greenwald density limit\supercite{Greenwald_density}. Between these bounds, $n_{85}$ was varied in PRO-create and passed through the STEP workflow to find temperature profiles that converged in TGYRO. The scale factor between electron and krypton density was scanned to obtain the required radiated power $P_{\rm rad}$, as calculated by TGYRO, such that scrape-off-layer power $P_{\text{SOL}}$ remained below 40 MW. 

While the core temperature was evolved by the transport calculations, $T_{85}$ was fixed during a TGYRO run but could significantly affect performance. The upper limit for $T_{85}$ was also set by maintaining ballooning stability of the edge pressure gradient (Fig. \ref{fig:BALOO_plot}). The lower limit for $T_{85}$ is that of a typical PT L-mode, but DIII-D NT plasmas have exhibited a steeper edge than PT L-mode plasmas while remaining ELM-free at sufficiently negative triangularity. The NT edge was therefore expected to lie somewhere between a typical L-mode and H-mode edge\supercite{Nelson_2023}. While enforcing H$_{98y2} < 1$ and an edge gradient within ballooning stability limits, $T_{85}$ was increased until TGYRO converged with $P_\text{fus} > 400$ MW, the estimated minimum value for MANTA to provide reasonable extrapolation to a commercial power plant (see Section \ref{sec:econ}).

To extrapolate the temperature and density profiles from the edge of TGYRO's evolution boundary ($\rho_{\rm tor}$ = 0.85) to the separatrix, PRO-create's modified $\tanh$ H-mode profile model was used to generate short pedestals that match closely those seen in DIII-D NT discharges. The pedestal width was set to be 0.1 units wide in normalized minor radius, consistent with more L-mode-like ELM-free operation. BALOO \supercite{MillerBALOO} evaluated the stability of this interpolated edge to verify the pressure gradient remained in the 1$^{\text{st}}$ ballooning stability region, as expected for a NT edge \supercite{Nelson_2022}. This is shown in Fig. \ref{fig:BALOO_plot}, where the normalized pressure gradients remain below the 1st stability limit for infinite-n ballooning modes, which is consistent with ELM-free operation.

\subsubsection{Transport simulation results} \hfill

The core scenario resulting from the integrated modeling workflow described above produced a plasma gain $Q_\text{p}=11.5$, $P_\mathrm{fus}=450\,$MW, $f_\text{Gr}=0.88$, $H_\text{98y2}$=0.79, and a $P_\text{SOL}$ of only 23.5 MW. This value of $P_\text{SOL}$ corresponds to a radiated power fraction of 0.82. As discussed in \cite{ARCH_2022}, this value is expected to be acceptable for an L-mode-like device. A complete list of MANTA's parameters is given in Table \ref{tab:MANTA_params}.

\begin{figure}
\centering
\includegraphics[width=\columnwidth]{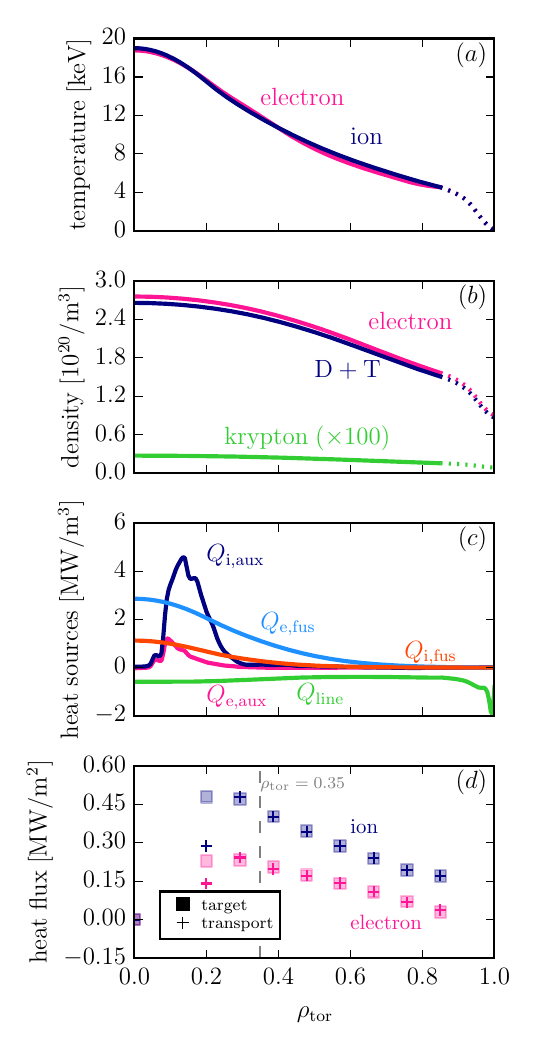}
\caption{Select TGYRO output profiles versus $\rho_\text{tor}$: (a) Electron and ion temperature profiles; (b) Electron, D + T, and krypton (scaled 100$\times$ for visibility) density profiles; (c) Heat deposition profiles for electron auxiliary heating $Q_\text{e,heat}$, ion auxiliary heating $Q_\text{i,heat}$, line radiation primarily from krypton radiation $Q_\text{line}$, electron $\alpha$ heating $Q_\text{e,fus}$, and ion $\alpha$ heating $Q_\text{i,fus}$; (d) Convergence of electron and ion heat flux profiles.}
\label{fig:TGYROprofiles}
\end{figure}

The temperature and density profiles are shown in Fig. \ref{fig:TGYROprofiles} (a) and (b). The heating sources for electrons and ions are illustrated in Fig. \ref{fig:TGYROprofiles} (c). The heat fluxes are well converged from $\rho_{\rm tor} = 0.35$ to $\rho_{\rm tor} = 0.85$ as seen in Fig. \ref{fig:TGYROprofiles} (d). Flux matching between $\rho_{\rm tor} = 0.0$ and $\rho_{\rm tor} = 0.35$ is known to be difficult, but it has been shown to have a marginal effect on output fusion power due in part to the relatively small plasma volume when compared to the edge \supercite{rodriguezfernandez2023}.

\section{Divertor and Power Handling}
\label{sec:divertor}

For the core scenario to be viable, heat fluxes on the divertor and first wall must be tolerable. These heat fluxes come from two main sources. The first is power transported through the SOL, $P_\mathrm{SOL}$, which streams along the open field lines to the divertor targets. The second is power radiated by the plasma via photons, $P_\text{rad}$, which more uniformly loads the vacuum vessel. Due to the small plasma-wetted surface area, the divertor targets are at greater risk of failure due to these heat fluxes.

Divertor research has thus focused on methods for mitigating the power fluxes incident on the divertor target plasma facing components (PFCs). This is especially important as parallel heat fluxes in the SOL are expected to increase by an order of magnitude for fusion pilot plants (FPP) compared to current devices \supercite{kuang_divertor_2020, menard_fusion_2022}. One mitigation method is the use of advanced divertor configurations \supercite{labombard_adx_2015, Kembleton_DEMO_2022, Wigram_ARCdivertor_2019} that feature flux expansion to spread out the heat flux or multiple X-points to distribute the heat flux onto additional targets. However, MANTA's radiative and NT operation permits a much simpler, conventional divertor to meet the reactor power exhaust challenge.

NT moves the X-points, and hence the entire divertor, to larger major radius. This effect is shown in Fig. \ref{fig:pt_vs_nt}. The larger circumference increases the area of the divertor targets relative to a PT divertor, spreading out the heat flux.  Additionally, ELM-free NT has no lower limit on $P_\mathrm{SOL}$, permitting higher levels of impurity seeding than would otherwise be possible. More power may thus be radiated away in the core prior to reaching the SOL, reducing $P_\mathrm{SOL}$. 

To quantify MANTA's divertor challenges, the metrics 
\begin{align}
    M_1 &= P_\mathrm{SOL}B_{T}/{R}\\
    M_2 &= (P_\mathrm{SOL}B_{T}/R)/n_\mathrm{sep}^{2}
\end{align}
 are considered for a variety of reactor-class devices. Here, $B_{T}$ is the toroidal magnetic field, $R$ is the major radius, and $n_\mathrm{sep}$ is the density at the separatrix. $M_1$ is a simple estimate of the relative parallel heat flux density, $q_{||} \propto P_\mathrm{SOL}B_{T}/{R}$ (assuming $\lambda_q \propto B_{p}$ \supercite{Eich_Hmode_2013}, where $\lambda_q$ is the SOL heat flux width, characterized as the e-folding length of $q_{||}$). $M_2$ includes the ability of a divertor to dissipate this power flux (since dissipation processes scale according to $\propto n_\mathrm{sep}^{2}$, similar to the Lengyel model \supercite{Moulton_Lengyel_2021}). The values of these metrics for MANTA, ARC V1 \supercite{Sorbom_APS_2020}, CFETR \supercite{Zhuang_CFETR_2019}, and EU-DEMO \supercite{Reimerdes_DEMO_2020} are listed in Table \ref{tab:divertorMetrics}. MANTA's divertor clearly operates in a far less challenging environment than that of other reactor-class tokamaks. This is a direct result of MANTA's ability to maintain a low $P_\mathrm{SOL}$ and high $n_\mathrm{sep}$, which is high relative to the core density as a result of smaller edge density gradients in ELM-free NT.

\begin{table*}[h!]
\begin{minipage}{\textwidth} 
\begin{center}
\caption{Comparison of divertor metrics between MANTA, ARC V1 \supercite{Sorbom_APS_2020}, CFETR \supercite{Zhuang_CFETR_2019}, and EU-DEMO \supercite{Reimerdes_DEMO_2020}}
\label{tab:divertorMetrics}
\begin{tabular}{llllll}
 \hline
 Parameter & MANTA & ARC V1 & CFETR & EU-DEMO \\
 \hline
 \hline
 $P_\mathrm{fus}$ (MW) & 451 & 500 & 558 & 2000 \\
 $R$ (m) & 4.55 & 3.65 & 7.2 & 8.8  \\
 $B_T$ (T) & 11.1 & 11.6 & 6.5 & 5.8 & \\
 $P_\mathrm{SOL}$ (MW) & 23.5 & 83 & 91 & 150 \\
 $n_\mathrm{sep} (10^{20} m^{-3})$ & 0.9 & 0.61\footnote{ARC $n_\mathrm{sep}$ estimated by assuming $n_\mathrm{sep}\sim$0.35*$\langle n \rangle$ for H-mode, taking $\langle n \rangle$ from \cite{Sorbom_APS_2020}} & 0.25 & 0.25 \\
 \hdashline
 $P_\mathrm{SOL}B_{T}/{R}$ & 57.3 & 263 & 82.2 & 98.9 \\
 $(P_\mathrm{SOL}B_{T}/R)/n_\mathrm{sep}^{2}$ & 70.7 & 707 & 1310 & 1580 \\
 \hline
\end{tabular}
\end{center}
\end{minipage}
\end{table*}

\subsection{Optimization of Poloidal Field Coils}

Starting from the CHEASE equilibrium, the free-boundary Grad-Shafranov solver FreeGS\supercite{FreeGS} was used to optimize the poloidal field (PF) coil set and finalize the equilibrium. The goal was to produce the simplest divertor geometry capable of tolerating $P_\mathrm{SOL} \approx 25$~{ MW} while maintaining adequate coil lifetimes. This resulted in three pairs of coils with currents similar to other ARC-class designs \supercite{kuang_divertor_2020, ARCH_2022}. Due to the demountability of the toroidal field (TF) coils, the PF coils are placed inside the TFs, reducing their size, current, and cost. 

\begin{figure}[h]
\centering
\includegraphics[width=\columnwidth]{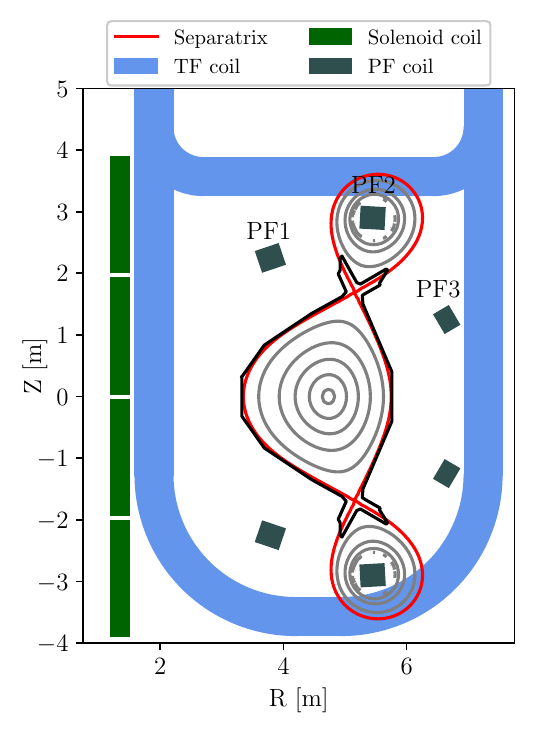}
\caption{Poloidal cross-section of MANTA showing the full set of REBCO-based superconducting coils and the resulting plasma equilibrium at maximum central solenoid current.}
\label{fig:freeGS_equi}
\end{figure}

Using the genetic algorithm described in \cite{DEBOUCAUD2024114401}, the optimal coil locations were determined by evaluating coil lifetimes (via an approximation of the full neutronics model described in Sec.~\ref{sec:neutronics}), coil areas, and coil currents for thousands of different possible locations. The coil rotation was also allowed to vary, resulting in the flat face of the coils being presented towards the core as this increased the amount of FLiBe shielding. Additionally, because the solenoid current changes throughout a pulse, this optimization evaluated each coil set at the most positive and most negative solenoid current, resulting in a coil set capable of performing well throughout a pulse. The optimal coil set and resulting equilibrium is presented in Fig. \ref{fig:freeGS_equi} at the time slice of maximum solenoid current. The current, size, locations, and additional details of the PFs are given in Table \ref{tab:PF_params}. The engineering considerations of the PFs are detailed in Section \ref{sec:magnets}. The PF lifetimes are given in Section \ref{sec:neutronics}.

\subsection{UEDGE Simulations of Scrape-Off Layer (SOL) Plasma}

Using the divertor geometry generated by FreeGS, the heat and particle flux along the open field lines to the divertor target plates were calculated with the 2D edge transport code UEDGE \supercite{rognlien_fully_1992}. UEDGE solves the Braginskii fluid equations for magnetized plasmas in the tokamak SOL to determine the steady-state plasma density, temperature, power and particle flows. It also utilises a fluid model to include computation of neutral dynamics and particle sources over the simulation domain. UEDGE has been previously applied to study divertors in high-field reactor-class devices \supercite{Wigram_ARCdivertor_2019, Ballinger_SPARC_2021} and is here used to refine the point-design and quantitatively assess divertor and power exhaust performance.

The simulation grid was created using the built-in UEDGE mesh generator. The resolution of the grid was increased near the separatrix and near the targets to resolve the steep plasma gradients and neutral recycling physics. At the outer midplane (OMP) separatrix, the grid resolution was chosen to be on the order of the gradient lengths associated with the parallel heat flux. Since MANTA operates in a double-null configuration, only the bottom half of the SOL was simulated to reduce computational cost. A non-orthogonal grid was used to allow fine-tuning of the target geometry relative to that of the field lines as imposed by the FreeGS equilibrium. The field line grazing angle on the divertor targets is ${\sim}3.25\degree$. Smaller angles were prevented by mesh cell distortion.

An input power of $P_\mathrm{SOL} = 25$ MW was applied at the radially innermost boundary of the UEDGE simulation domain, divided evenly between ions and electrons for simplicity. This is marginally higher than the 23.5 MW obtained from transport simulations in Section \ref{sec:core} due to numerical difficulties related to detachment preventing a smaller $P_\mathrm{SOL}$ case from converging. These same numerical difficulties forced a separatrix density $n_\mathrm{sep}$ of $0.85 \times 10^{20}$ $m^{-3}$, rather than the $0.9 \times 10^{20}$ $m^{-3}$ value used in the core modelling. 
Between the larger tilt angles, higher $P_\mathrm{SOL}$, and lower $n_\mathrm{sep}$, the results presented here are a conservative estimate for the divertor performance, and the divertor heat flux is likely lower than that calculated here.

Extrinsic impurity seeding \supercite{loarte_plasma_1998, leonard_scaling_2012, kallenbach_impurity_2013} in the SOL was required to reduce the power incident on the divertor targets. Neon was selected as the radiator for its low atomic number, ability to radiate effectively at SOL temperatures, and the inertness and recycling benefits of noble gases. Impurity seeding was included in UEDGE via a fixed-fraction impurity model, where the impurity density $n_\mathrm{imp}$ was chosen as a fraction of the main ion density, $f_\mathrm{imp} = n_{\rm imp}/n$. An additional benefit of not operating in H-mode is the lack of a edge transport barrier; accumulation of impurities in the core is not expected and thus seeding of impurities in the SOL is unlikely to be problematic.\supercite{Sciortino_2022}


Transport coefficients are chosen to match expected scalings for the heat flux width and in-out power asymmetries. Multi-machine scalings for type-I ELMy H-modes \supercite{eich_empiricial_2013} and high-field specific scalings across confinement regimes \supercite{brunner_high-resolution_2018} indicated that for MANTA's $B_{p}$, $\lambda_{q}$ could be as small as $0.3$\,mm. On the other hand, a PT L-mode specific scaling \supercite{horacek_scaling_2020} predicted $\lambda_q$ of at least a few mm. Recent work on DIII-D and TCV in NT places $\lambda_{q}$ between $\lambda_{q}$ for H-mode and PT L-mode\supercite{faitsch_dependence_2018, Scotti2024}. As a compromise between the different scalings, the transport model was chosen to give $\lambda_{q} = 0.9 $ mm, as measured using an exponential fit to $q_{||}$ at the divertor throat. In order to reach small enough $\lambda_{q}$ predicted by the $B_{p}$ scaling, the ion and electron heat diffusivity, $\chi_{i,e}$, was reduced to below $10^{-2}$ m$^2$s$^{-1}$. 

Evidence exists for ballooning-like transport on the LFS \supercite{labombard_particle_2001}, implying a larger flow of plasma to the outer than inner divertor. Double-null configurations are known to produce larger in-out asymmetries ($>$80\% power exhausted on the LFS) due to the HFS and LFS SOLs being magnetically disconnected \supercite{Brunner_powersharing_2018, DeTemmerman_asymmetries_2010}. The power split may not be as asymmetric in NT due to the decreased volume on the ``bad curvature" side when compared to PT, resulting in reduced interchange-driven turbulence on the LFS \supercite{lim_turbulence_2023}. MANTA therefore assumed a 70/30 power split between the outer and inner divertors respectively. This is done by increasing $\chi_{e,i}$ at the outboard side of the simulation grid relative to its value at the inboard side. The particle diffusivity, $D$, was chosen such that density gradient scale length $\lambda_{n} = \frac{n}{\nabla n} \approx 6-10$ mm, as expected at the separatrix in L-mode-like plasmas \supercite{ballinger_edgeprofiles_2022,silvagni_edgeprofiles_2020}. Since an H-mode like transport barrier and resulting pedestal is not expected in ELM-free NT operation, flat transport profiles are used in the radial direction. 

Though $^4$He is not explicitly modeled in UEDGE, a pumping condition is still applied in the domain to the main ion species to account for $^4$He removal and the impact on divertor neutral gas pressures, plasma flows, etc. Assuming that 2\% of the plasma exhausted from the core is $^4$He, and an enrichment of $\sim$0.75 in the divertor  (consistent with previous studies\supercite{Hillis_enrichment_1999}), the pumped main ion flux required to be removed in steady-state is found to be $\sim10^{22}$ particles/s. This condition is applied as a fixed particle flux removed from the simulation domain across the UEDGE private flux region boundary cells in the region identified in Fig. \ref{fig:uedge_up} (a), placed close to the outer target where neutral gas densities are highest for efficient pumping. The values for the physics parameters chosen for the UEDGE simulation are tabulated in table \ref{tab:uedge_params}.

\begin{table}[H]
\begin{center}
\caption{Principal control parameters used in UEDGE simulation.}
\label{tab:uedge_params}
\begin{tabular}{lll}
\hline
Parameter & Value & Units \\
\hline
\hline
$P_\mathrm{SOL}$ & 25 & MW \\
$n_\mathrm{sep}$ & 0.85 & $\mathrm{10^{20} m^{-3}}$ \\
$f_{\rm Ne}$ & 0.315 & $\mathrm{\%}$ \\
$\Gamma_{\rm pump}$ & 5 $\times 10^{21}$ & $\mathrm{s^{-1}}$ \\
$\chi_{\rm i/e, out}$ & $1.5 \times 10^{-2}$ & $\mathrm{m^{2}s^{-1}}$ \\
$\chi_{\rm i/e, in}$ & $4 \times 10^{-3}$ & $\mathrm{m^{2}s^{-1}}$ \\
$D$ & $2.5 \times 10^{-2}$ & $\mathrm{m^{2}s^{-1}}$ \\
\hline
\end{tabular}
\end{center}
\end{table}

\subsubsection{Edge Modeling Results}

\begin{figure}[h!]
\centering
\includegraphics[width=\columnwidth]{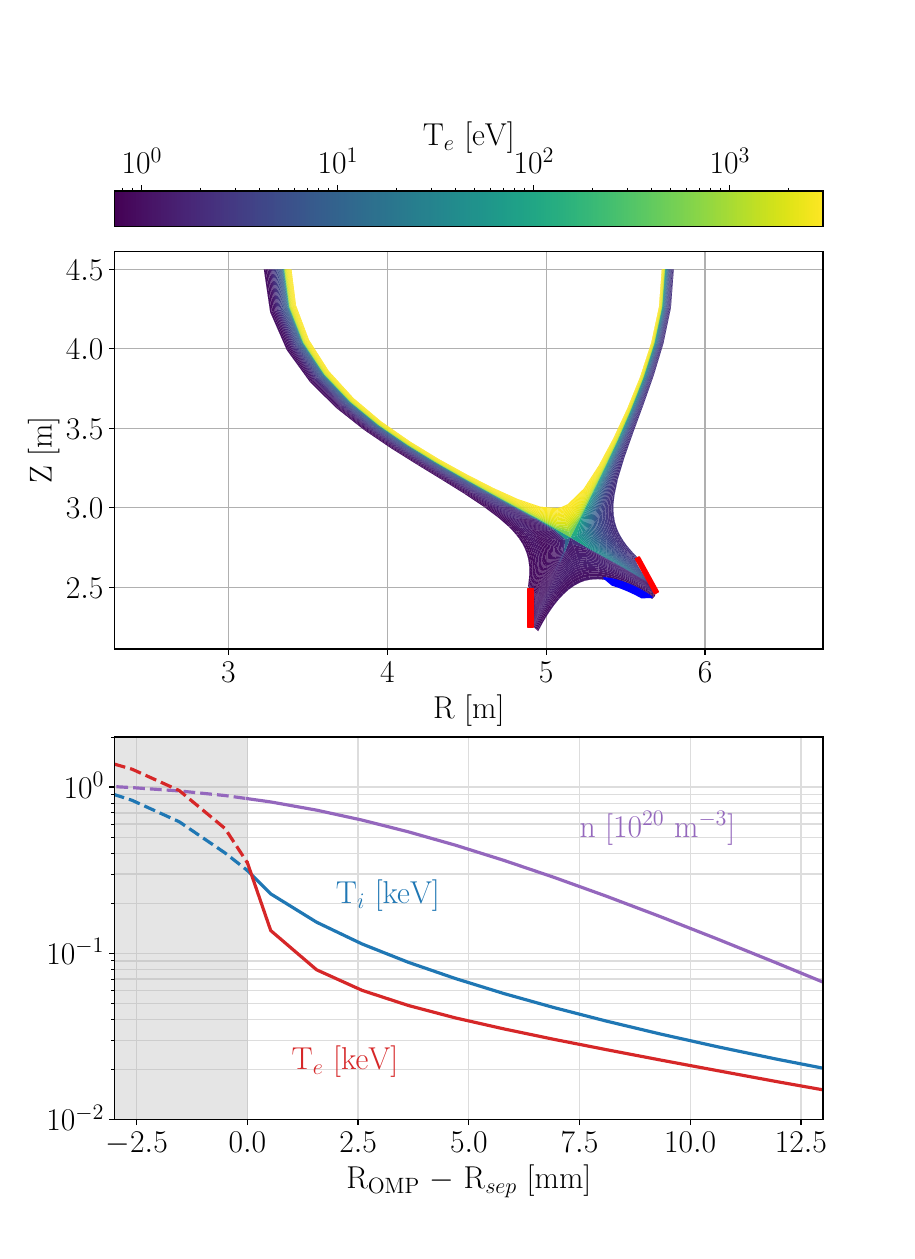}
\caption{(a) 2D contour of electron temperature ($T_{e}$) for the final UEDGE solution. The outer and inner targets are shown in red and the pumping surface is shown in blue. Note the much larger temperatures in the outer SOL than in the inner SOL. (b) Upstream profiles of $n$ (purple), $T_{i}$ (blue), and $T_{e}$ (red) at the outer midplane (OMP). The gradient scale lengths at the separatrix are smaller for the temperature than the density}
\label{fig:uedge_up}
\end{figure}

The results of the UEDGE simulations are shown in Fig. \ref{fig:uedge_up}. Plotted in (a) is $T_{e}$ in the lower half of MANTA's SOL, where the inner leg and more generally the HFS are significantly colder than the outer leg and LFS due to the prescribed 70/30 outer/inner power split. There are no up-down asymmetries present in the simulation as drifts were not included. Fig. \ref{fig:uedge_up} (b) shows upstream $n$, $T_{e}$, and $T_{i}$ profiles at the outer midplane. 

\begin{figure}[h!]
\centering
\includegraphics[width=\columnwidth]{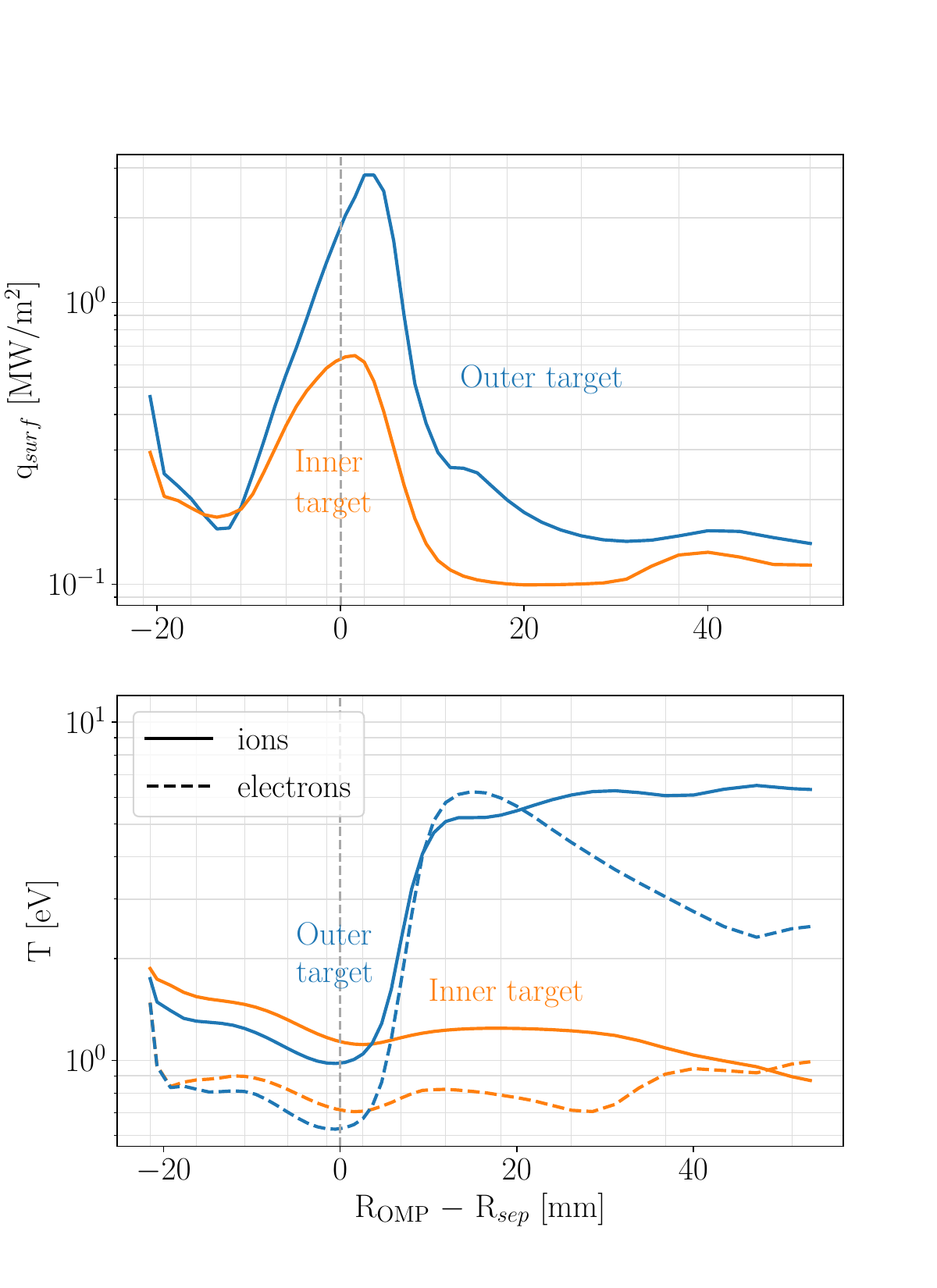}
\caption{(a) Heat flux density arriving at the inner (orange) and outer (blue) divertor targets. The heat flux density is much larger on the outer target, expected from the in-out power asymmetry modeled. (b) $T_{i}$ (solid line) and $T_{e}$ (dashed line) on inner target (orange) and outer target (blue). Temperatures near the strike point are well below 5 eV on both targets. $T_e$ and $T_i$ peak above above 5 eV at least $15$ mm from the strike point.}
\label{fig:uedge_down}
\end{figure}

Figures \ref{fig:uedge_down} (a) and (b) illustrate the divertor heat flux and temperature profiles, which ultimately dictate divertor survivability. The inner target is fully detached and the outer target is partially detached. With an impurity fraction of just 0.315\%, the more heavily loaded outer target has a peak heat flux of only 2.8 MW/m$^{2}$, well under the usually quoted heat flux limit of 10 MW/m$^{2}$ \supercite{hirai_iter_2014}. However, this target experiences a maximum $T_{e} = 6.3$\,eV and $T_{i} = 6.5$\,eV, slightly higher than the desired $5$\,eV limit to minimize W sputtering \supercite{brooks_analysis_2013}. These maxima occur $15$\,mm and $47$\,mm from the strike point, much farther into the SOL than the peak of the heat flux profile, ($\sim2$\,mm from the strike point), where the particle flux is much lower and therefore may not present a large concern.

To evaluate the level of W erosion, sputtering dynamics under the present target conditions are analysed for the D, T, He and Ne species. At conditions for $T_{e} < 10$\,eV, sputtering contribution from D-T ions is negligible\supercite{Brezinsek_sputtering_2019}, such that sputtering is primarily driven by impurities. The impurity ion impact energies (taking $E_{in} = 3ZT_e + 2T_i$, where Z is the average charge state\supercite{Brezinsek_sputtering_2019}) are evaluated and related to the relevant sputtering yield $Y_W$ data. At $T_e\sim5$\,eV, $E_{in}$ is below the sputtering energy threshold for helium (117 eV)\supercite{Hwangbo_heliumsputt_2017}, hence no sputtering contribution is anticipated for the He species. Neon sputtering yield is evaluated at $Y_W\sim10^{-4}$ for $E_{in} = 40$\,eV \supercite{Meluzova_neonsputt_2020,Zhao_erosion_2020,Zhao_erosion_2020}. Main ion particle fluxes around the target $T_e$ peak are at maximum $\Gamma_D \sim 1\times10^{22}$\,m$^{-2}$s$^{-1}$. Combining $Y_W$ and $\Gamma_D$ with Equation (3) in \cite{Zhao_erosion_2020}, the W sputtering erosion rate from 0.315\% Ne fraction is estimated at 0.0016 mm/year - therefore having no impact on the divertor lifetime. 

\subsection{Divertor and Vacuum Vessel Heat Removal}

\begin{figure}[h]
\centering
\includegraphics[width=1\columnwidth]{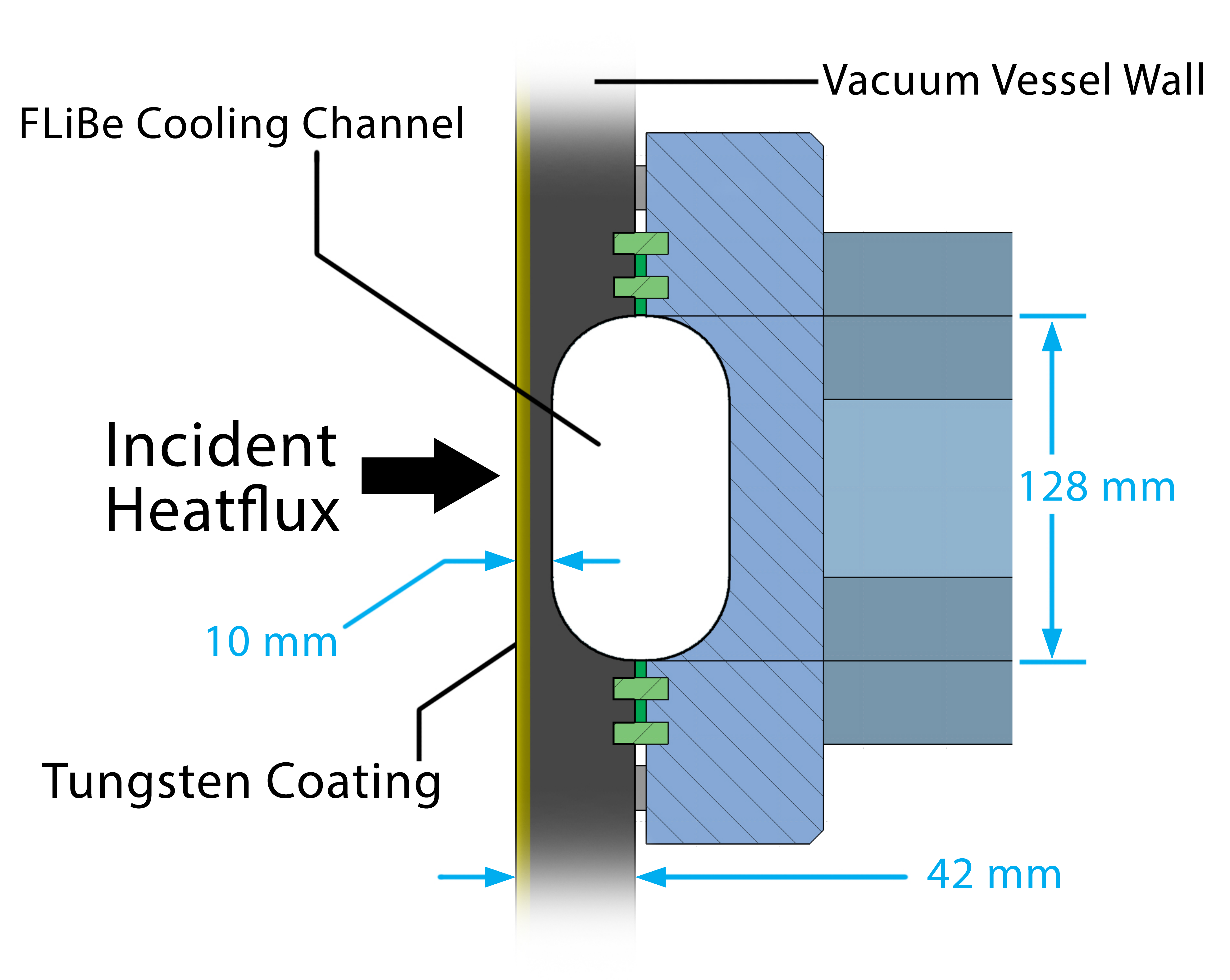}
\caption{Poloidal cross section of a VV-wall divertor target with a FLiBe cooling channel. The tungsten plasma-facing surface is indicated in gold, the alloy sealing gaskets are highlighted in green, and the external channel block is colored blue.}
\label{fig:divertorPanelCrossSection}
\end{figure}

Active cooling of the first wall components was required, and focus in this study was given to developing a scheme for cooling the divertor targets as they must withstand the highest local heat fluxes. This is achieved via FLiBe channels in direct contact with the backside of the targets, arranged in 18 toroidal segments around the VV. A poloidal cross section of a target is given in Fig. \ref{fig:divertorPanelCrossSection}.

\begin{figure}[h]
\centering
\includegraphics[width=1\columnwidth]{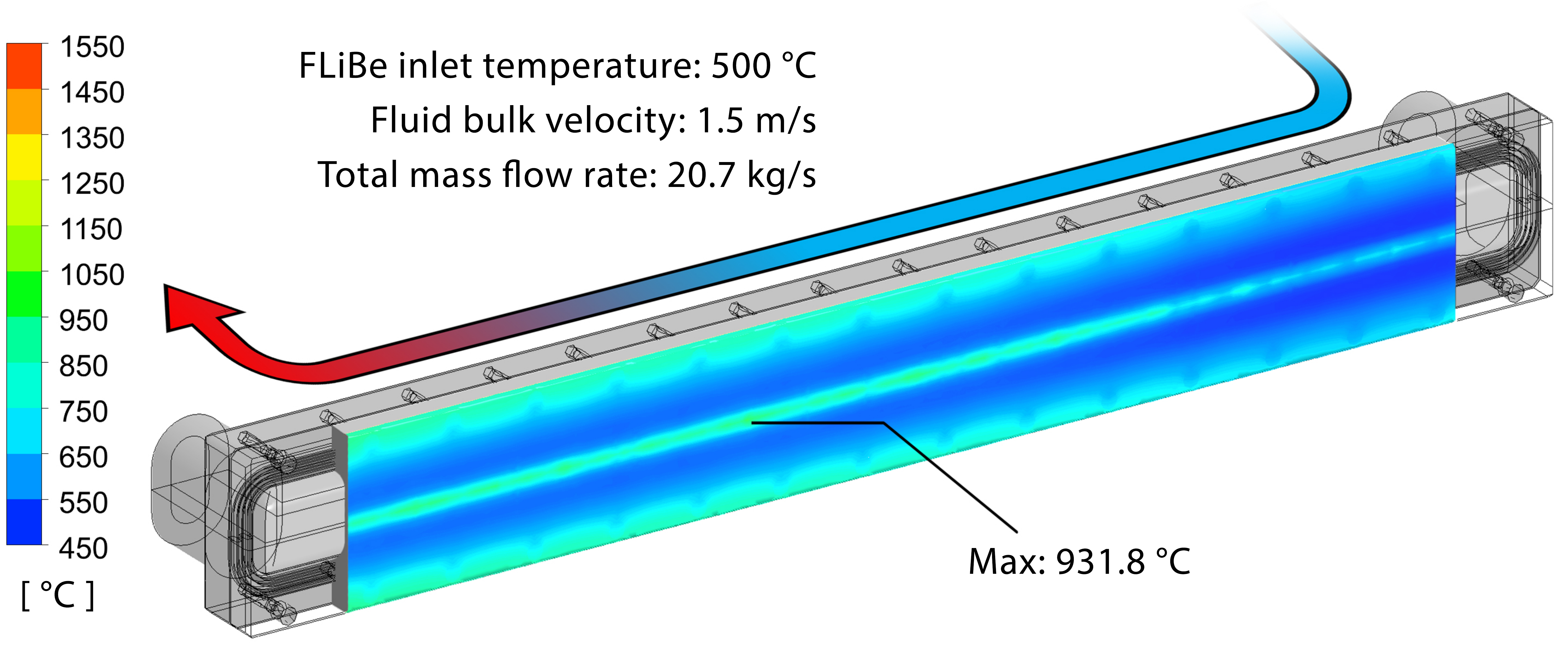}
\caption{Results of the CFD simulation for the outboard divertor target. Perpendicular inlets and outlets were included to improve fluid fidelity, and the thermal results span between those ports. The bulk temperature increases along the length of the panel. The peak temperature remains far below the Tungsten recrystalization temperature ($1550\degree C$).\supercite{deformedTung}}
\label{fig:ANSYS_temp}
\end{figure}

Ansys Fluent\supercite{ANSYS} was used to predict the temperature this target design would experience due to the heat flux calculated by UEDGE. The total mass flow rate of the FLiBe was taken to be 20.7 kg/s, resulting in a bulk velocity of 1.5 m/s. In these simulations, the VV wall was considered a single piece of tungsten to avoid simulation artefacts in the thin geometries. This simplified model also neglected toroidal curvature, fasteners, as well as any additional heat removal through the gasket interface. The temperature around the inlet and outlet were not included in the simulation as they were not optimized. The resulting target temperature is shown in Fig. \ref{fig:ANSYS_temp}, where the maximum temperature is only ${\sim}930\degree C$, well below the recrystalization temperature of tungsten ($1550\degree C$\supercite{deformedTung}).

A conformal vacuum vessel (VV) design was chosen to maximize the tritium breeding ratio (TBR) and magnet shielding. A gap of 10 $\lambda_q$, or about 1 cm between the separatrix and VV was chosen to ensure the vast majority of $P_\mathrm{SOL}$ enters the divertor region and does not interact with the VV. Active cooling of the VV was not fully investigated, but one potential solution is a double-walled VV in which FLiBe flows between the two shells and exhausts into blanket tank along with the divertor coolant.

\section{Magnet Design and Device Maintenance}
\label{sec:magnets}

MANTA's high magnetic field is produced by 18 non-insulated REBCO HTS toroidal field (TF) coils, a technology pioneered by the SPARC Toroidal Field Model Coil (TFMC) project \supercite{hartwig2023sparc, Rui_TFMC, Whyte_TFMC_valid}. The central solenoid (CS) and poloidal field (PF) coils are also made of REBCO tape, though the need for low AC losses and fast response required insulated magnets composed of PIT-VIPER-like cables \supercite{hartwig2020viper}. All magnets are operated at 20 K with liquid hydrogen (LH2) coolant. Additionally, MANTA was designed around maintaining a reasonable duration for its maintenance cycles, which is primarily a function of the magnet ramp times and cryostat thermal cycling.

\subsection{Toroidal Field Coils}
\label{sec:TF_magnets}

The TF magnet design was driven by the need to achieve 11 T on axis while minimizing the required length of REBCO tape, keeping stress within engineering tolerances, and maintaining enough FLiBe between the plasma and the coil for neutron shielding. Additionally, the TFs are designed to be demountable to allow for maintenance and replacement of internal components, namely the vacuum vessel (VV) and PF coils. Several shapes were evaluated for how well they could accommodate all of these requirements, including the traditional ``Princeton Dee'' \supercite{PrincetonDMagnet}, a reversed Dee, and ``window pane'' \supercite{CmodTFs}. 

\begin{figure}[h]
    \centering
    \includegraphics[width=0.8\columnwidth]{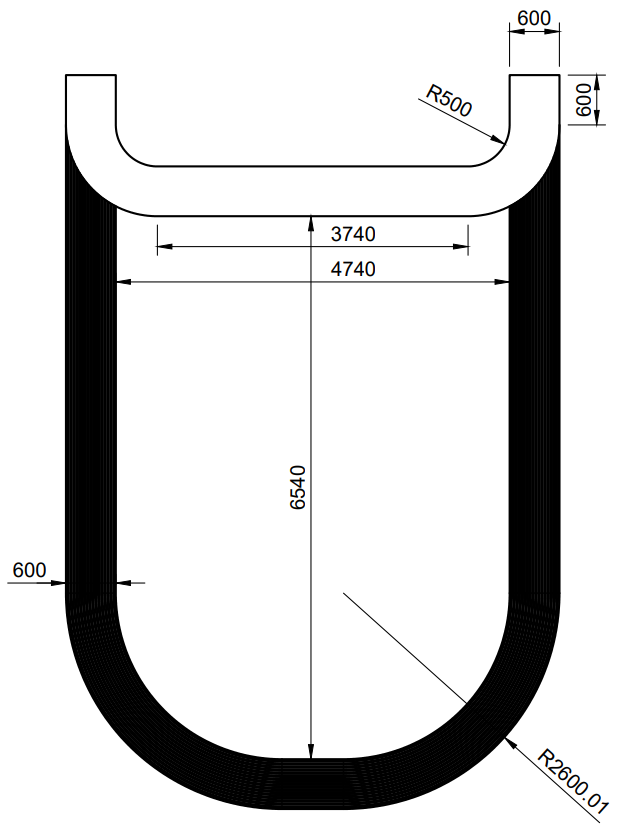}
    \caption{Schematic showing TF 2D planar dimensions in mm. Total TF azimuthal thickness is 544mm. Each of the 18 pancakes is 28mm thick. The magnet casing adds another 40mm to the azimuthal thickness, making the total azimuthal thickness of the TF 544mm.}
    \label{fig:vikinghorns}
\end{figure}

The final shape, shown in Fig \ref{fig:vikinghorns}, is a variation of the window pane design. Of the several shapes evaluated with COMSOL \supercite{COMSOL}, this design minimized peak stress while accommodating the joints. At full-field, the maximum von Mises stress is 600 MPa. This is acceptable for the chosen baseplate material, Inconel-718, which has a yield strength above 1000 MPa at cryogenic temperatures \supercite{tobler1976low}.

COMSOL simulations indicated $13.6$ MA-turns were required to produce the necessary 11 T field on axis. To achieve this, each TF is made up of 18 pancakes with 16 turns per pancake, giving an operating current $I_{\rm op}$ of $47.2$ kA. The REBCO tape stack was designed with a height $h$ of $4$ mm and a width $w$ of 21 mm, resulting in an operating current density $J_{\rm op}$ of $570\, \text{A}/\text{mm}^2$. This gives a $40\%$ margin below the 25 K and 25 T critical current density of $J_{c}=1000\,\text{A}/\text{mm}^2$. This $\rm J_{c}$ is obtained from recent experimental data from commercially available superOx tapes with magnetic fields up to 30 T\supercite{Molodyk2021}. The magnetic field in this experiment was aligned perpendicular to the plane of the tapes, a worst case scenario for $J_{c}$.

\begin{figure}[h]
    \centering
    \includegraphics[width=\columnwidth]{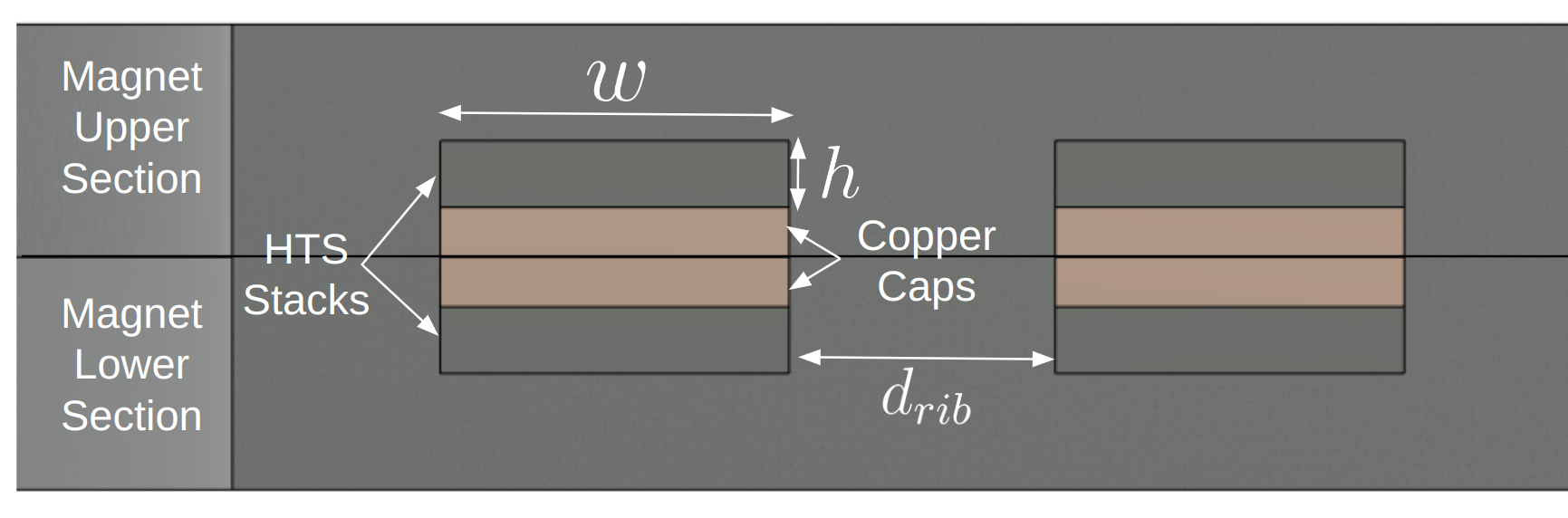}
    \caption{Cross-sectional view of the magnet at the joint, where the upper and lower sections meet. }
    \label{fig:tf_cross}
\end{figure}

While a detailed quench resilience analysis is outside the scope of this paper, given the evolving nature of the field, quench resilience needs are anticipated by making room for a $3\times 21$ mm copper cap, a leading approach to improving quench resilience of the magnet \supercite{mouratidis2024performance}. To ensure variations in the radial current between magnets do not cause unacceptable toroidal ripple fields, the approach detailed in \cite{mouratidis2022low} is used, where rib thicknesses and nominal joint resistances are designed to be low enough such that the radial current fraction, $I_r$/$I_{\rm op}<0.5$\%. This is done by setting: 
\begin{align}
    d_{\rm rib}\geq \frac{2 R_j h \cdot \text{max}(l_{\rm up}, l_{\rm low})}{\rho_p}\left(\frac{I_{\rm op}}{I_r} - 1\right),
\end{align}
where $d_{\rm rib}$ is the rib thickness, $R_j$ is the resistance per joint, $h$ is the height of each tape stack, and $\rho_p$ is the baseplate resistivity. Fig. \ref{fig:tf_cross} provides a cross-sectional view of the TF showing how key elements fit within the geometrical constraints of the pancake.

\begin{table*}[h]
\caption{TF Coil Parameters}
\resizebox{\textwidth}{!}{
\begin{tabular}{ llll||llll }
 \hline
 Symbol & Parameter & Value & Units & Symbol & Parameter & Value & Units \\
 \hline
 \hline
$N_{\rm tf}$ & Number of TFs & 18 & - &  $I_{\rm op}$ & Operating current & 47.2 & kA\\
 $N_{\rm pan}$ & Pancakes per TF & 18 & - &$J_{\rm op}$ & Operating current density & 570 & $\text{A/mm}^2$\\
 $N_{\rm tpp}$ & Turns per pancake & 16 & - &$J_{\rm crit}$& Critical current density & 1000 & $\text{A/mm}^2$\\
  $L$ & Inductance & 5.62 & H&$R_{\rm rpt}$& Radial resistance per turn & 150 & n$\Omega$\\
$W_{\rm TF}$& Stored energy & 6.27 & GJ&$R_c$& Characteristic resistance & 40.4 & $\mu\Omega$\\
$d_{\rm rib}$& Rib thickness & 0.016 & m&$\rho_p$& Baseplate resistivity & 0.982 & $\mu \Omega$-m\\
 $l_a$ & Azimuthal thickness & 0.544 & m& $R_j$& Resistance per joint & 0.5 & n$\Omega$\\
 $l_r$ & Radial thickness & 0.6 & m &$w$& Width of each REBCO stack & 0.021 & m\\
 $l_{\rm up}$ & Upper mean length & 7.455 & m&$h$& Height of each REBCO stack & 0.004 & m\\
 $l_{\rm low}$ & Lower mean length & 19.2 & m&$h_{\rm cu}$& Height of the copper cap & 0.003 & m\\

 \hline
\end{tabular}}
\end{table*}

\subsection{Maintenance}
\label{ssec:magnets/maintenance}
Maintenance in a reactor-class tokamak is expected to be extremely time-consuming due to the nuclear environment in the vessel, the limitations of remote maintenance, and the time required for the magnets to ramp-up/down and for the cryostat to cool/warm \supercite{Maris_2023}. Past ARC-class concepts have proposed to hasten maintenance cycles by demounting the top of the TF coils and replacing the VV wholesale \supercite{ARC_2015}. This scheme could allow for faster VV replacement than traditional approaches, which instead rely on removing heavy blanket modules through the gaps of the magnet cage \supercite{tesini2008iter}.

MANTA expands on the demountable magnet philosophy in two ways: 1) replacing the FLiBe tank, PF coils, and VV together (jointly referred to as the ``internal assembly'') and 2) oversizing the cryoplant relative to the cooling power requirements of nuclear operations. Removing the internal assembly in one piece eliminates the need to open the FLiBe tank during VV replacement, thereby significantly reducing the dose rate that equipment and workers would be exposed to (see Section \ref{subsec:component_activation} for more details). This enables safer and quicker FLiBe tank replacement.  

\begin{figure}[h]
    \centering
    \includegraphics[width=\columnwidth]{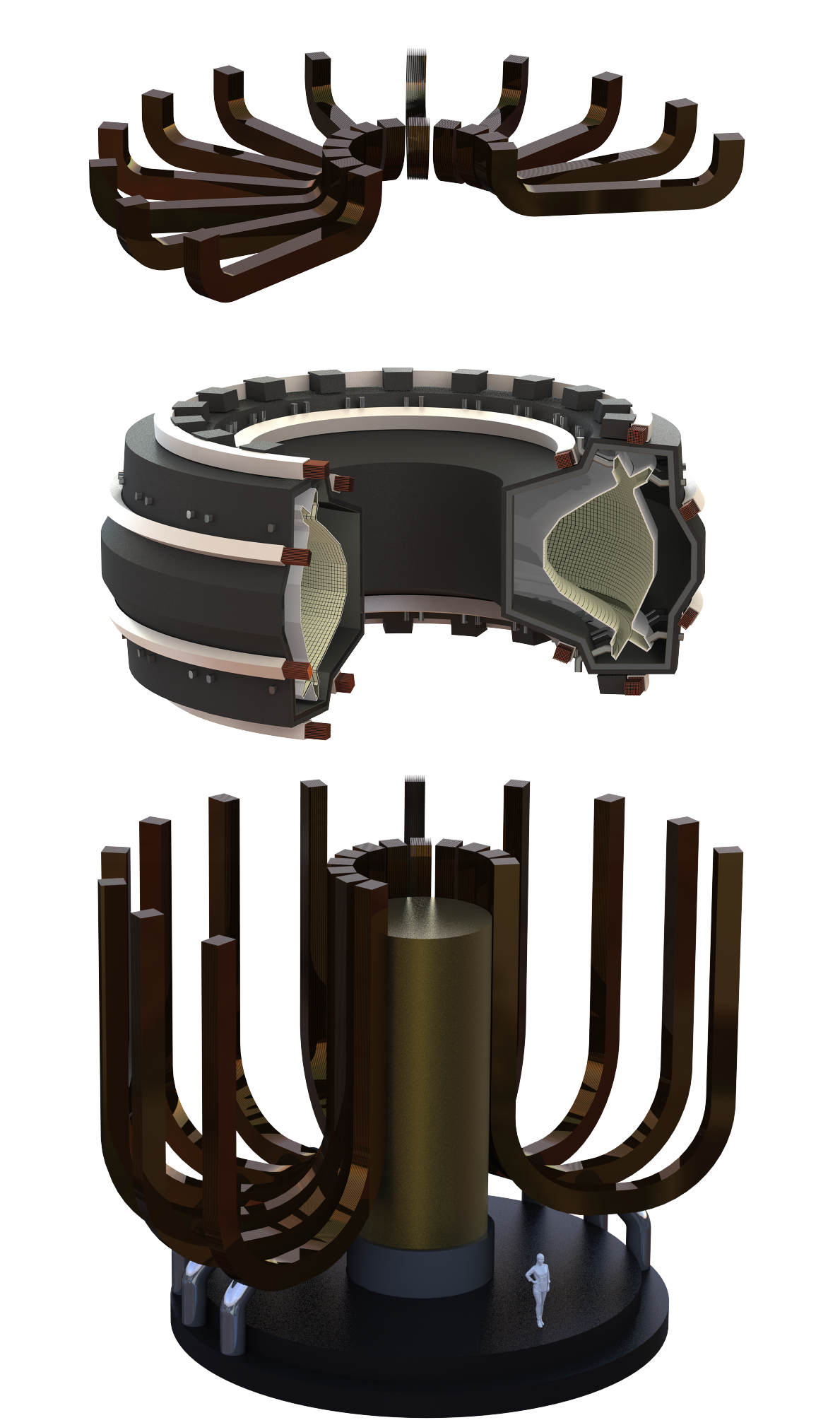}
    \caption{During maintenance, the TFs are demounted, and the VV, FLiBe tank, and PFs are extracted together as a single unit.}
    \label{fig:maintenance}
\end{figure}

The oversized cryoplant is key to reducing the two rate-limiting maintenance steps: TF magnet charging and temperature cycle time. Traditional insulated superconducting magnets can be ramped quickly by driving high voltages, but doing so in non-insulated magnets drives radial currents, which heat the magnet. Current-ramp rates in non-insulated magnets are therefore constrained by the cryogenic cooling capacity, which in practice causes non-insulated coils to take far longer to cycle. For example, KSTAR can fully ramp its insulated TF coils in ${\sim}$30 minutes \supercite{chu2011estimation}, while the SPARC Toroidal Field Model Coil (TFMC) required ${\sim}$20 hours to charge \supercite{hartwig2023sparc}. Equation \ref{eq:mag_ramp_limit}, derived in \ref{subsec:mag_ramp_deriv}, provides the following bound on the maximum magnet ramp-rate:
\begin{equation}\label{eq:mag_ramp_limit}
    \frac{dI_{\rm op}}{dt} \leq \frac{\sqrt{R_cP_{\rm ramp}}}{L},
\end{equation}
where $L$ is the magnet inductance, $R_c$ is the magnet characteristic resistance, and $P_{\rm ramp}$ is the cooling power available to a single TF to remove heat generated due to magnet ramping. The inverse dependence on inductance exacerbates the current-ramp-rate problem for large non-insulated magnets such as those on MANTA, which have 40x higher inductance than the SPARC TFMC \supercite{hartwig2023sparc}. A simple extrapolation from the SPARC TMFC with constant $R_c$ and $P_{\rm ramp}$ would stretch the charging time to over 67 days. $L$ is driven by the on-axis field requirements and system geometry and is thus not significantly adjustable. $R_c$ is driven by rib thicknesses which are ultimately limited by the magnet geometry. Thus, the primary adjustable variable to decrease the magnet ramp time is $P_{\rm ramp}$.

To achieve reasonable magnet charging times, $P_{\rm ramp}$ must far exceed the requirement for keeping the magnets at cryogenic temperatures during nuclear operations, $P_{\rm nuc}$. Neutronics calculations (Section \ref{sec:neutronics}) indicate that 2.1 kW of heat is deposited in the TFs by neutrons and an additional 0.20 kW in the PFs when MANTA is operating at a typical 450 MW. Owing to the excellent shielding provided by the FLiBe, this is small compared to the 11.5 kW of joint heating. Setting $P_{\rm ramp} = (13.8 \text{ kW}/18 \text{ TFs}) = 0.77$ kW/TF would result in a ramp-down time of 42.6 days. Taking into account the equally-time consuming ramp-up time and neglecting any repair time, a single unscheduled maintenance event per year would reduce the plant's availability by 22.3\% \supercite{Maris_2023}. This is clearly unacceptable for a power plant; for comparison, the global nuclear industry averages an Unplanned Capability Loss Factor of 3-6\% \supercite{noauthor_pris_nodate}. 

MANTA's cryosystem was therefore designed to provide 200 kW of cooling power during the magnet ramp phase, reducing the magnet cycle time to just 4.6 days. The cost of this oversized cooling power is more than offset by the higher availability factor this implies. Such a large cryoplant is also beneficial for quickening the temperature cycle of the cryostat (i.e. the time to go from cryogenic temperatures to room temperature, and back again). The temperature cycle of the KSTAR cyrostat lasts several weeks\supercite{chang2010operation}, and the temperature cycle for ITER will be similar \supercite{peng2013thermal}. To emphasize again, weeks-long delays before maintenance can begin will severely affect plant availability.

The energy required to increase the TFs from 20 K to 300 K was estimated by assuming there is approximately 1.1 million kg of Inconel-718 and 0.3 million kg of REBCO. REBCO's properties were approximated by a 44\%/56\% mix of copper and Hastelloy. The specific heat of Inconel and Hastelloy were approximated by that of stainless steel, similar to \cite{mangiarotti2013experimental}. Numerically integrating over empirical specific heat curves yielded an estimate of 140 GJ of warming/cooling energy required to raise/lower to complete a temperature cycle of the TFs. In order to achieve a week-long warming/cooling cycle time and take advantage of the availability and low-cost of liquid nitrogen (LN2), a two-step temperature ramp process was assumed. When temperatures of are above 77 K, LN2 is used as a precoolant for LH2, increasing the cooling power up to 500 kW. Below 77 K, 200 kW of cooling power is available, the same as during the magnet ramps. While this first step will take 3.0 days, the low specific heat of steel and copper at cyrogenic temperatures enable the second step to be completed in only 0.46 days. This results in a total temperature cycle time slightly less than one week.

\begin{figure}[h]
    \centering
    \includegraphics[width=1\columnwidth]{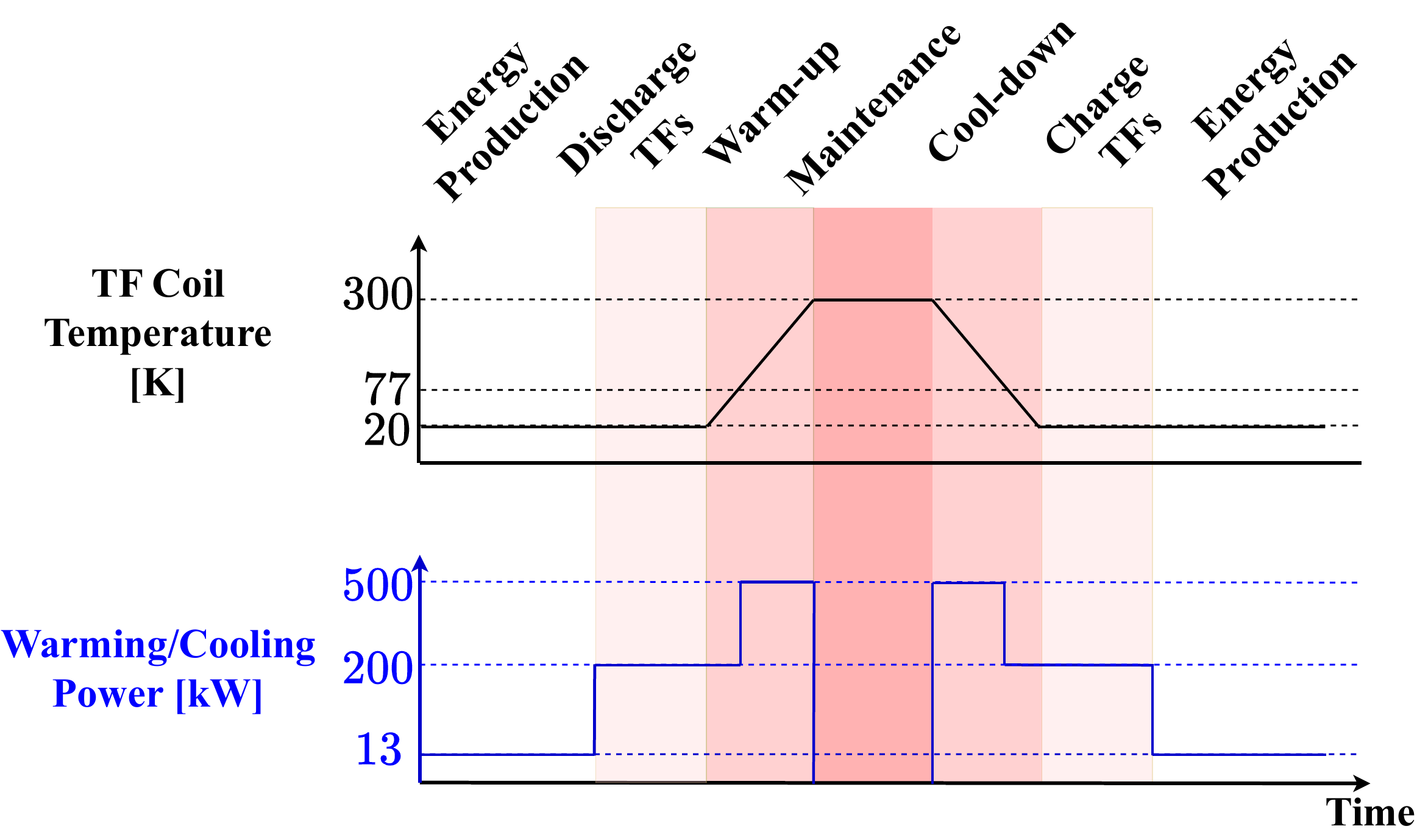}
    \caption{A diagram of the maintenance cycle in terms of temperature of the TFs and cooling/warming power.}\label{fig:cool_time_diagram}
\end{figure}

A schematic of the entire maintenance cycle is shown in Fig. \ref{fig:cool_time_diagram}. Cooling channels in the magnets must be designed to handle these non-standard cooling and heating powers over a wide temperature range. While not a trivial problem, these fast magnet and temperature cycle times are essential for a pilot plant (which may face unscheduled maintenance often) and for power plants (which must achieve high availability factors). A traditional ARC-class device with a 42.6 day magnet ramp time and a three-week long temperature ramp time would be offline for at least six months for any maintenance event, but MANTA with ``oversized'' cooling/heating power can complete both cycles in 16.1 days. 

\subsection{Central Solenoid and Poloidal Field Coils}
\label{sec:central_sol}

The mechanical stability of the Central Solenoid (CS) and Poloidal Field coils (PF) was investigated with a 2D-axisymmetric finite element model developed in COMSOL Multiphysics\supercite{COMSOL}.  The CS and PFs are modeled as homogenized multi-turn coils consisting of VIPER cables \supercite{hartwig2020viper} with four twisted tape stacks. The contribution of the plasma current to the overall magnetic field is approximated by a single-turn elliptical cross section coil with a uniform current profile. The geometry, materials and boundary conditions on this model are detailed in \ref{appx:cspf}.

To maximize the magnetic flux available to drive the plasma current while respecting engineering stress limits, a maximum operating current density of $J_{\rm op}$ = 80 A/mm$^2$ was selected. This choice affords a 25\% margin to the VIPER cable critical current at 20 K and 25 T. This current density, together with MANTA's 3.2 m inner bore diameter, allows the CS to provide 260 Wb of flux with a peak field of 25 T on the CS coils. Using the flux consumption approximations for startup from \cite{Sugihara_fluxes} and the loop voltage calculated from the equilibrium, 260 Wb produced a pulse length of ${\sim}15$ minutes.

\begin{figure}
    \centering
    \includegraphics[width=\columnwidth]{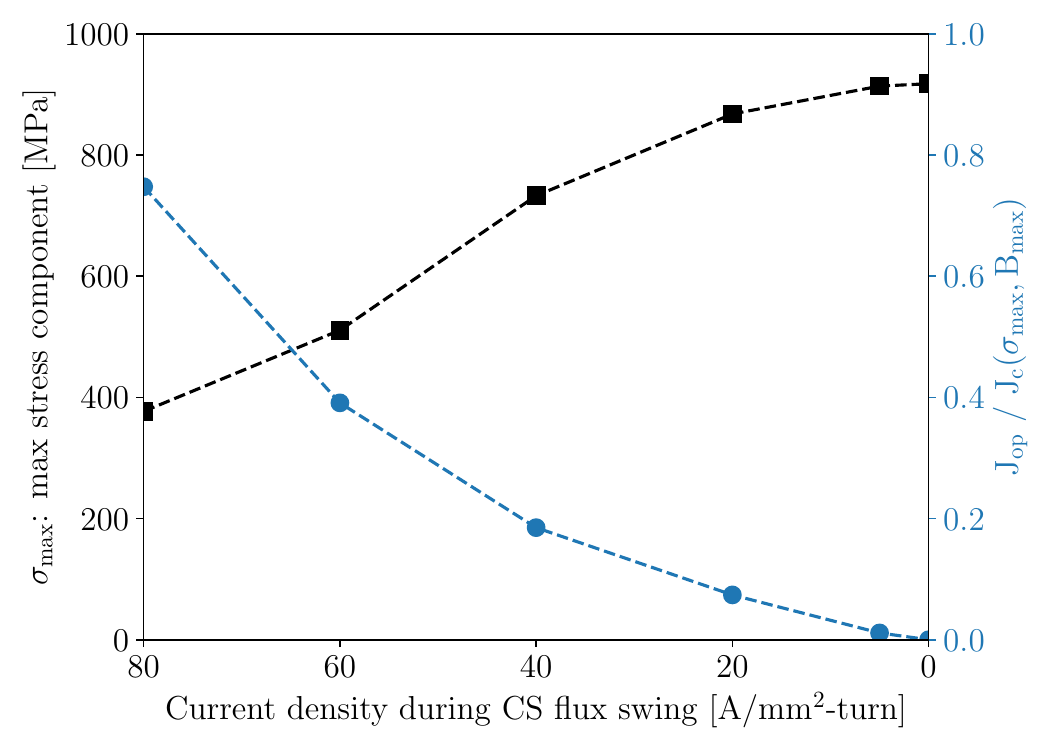}
    \caption{Throughout the current swing of the CS, the maximum stress component remains below the yield stress of Inconel ($\rm {\sim} 1000\,MPa$) and the critical current density of SuperOx tapes at 20 K and 25 T under electromagnetic stress loading.}
    \label{fig:jc-stress}
\end{figure}

The largest stress in the CS occurred at zero solenoid current $I_\text{CS}$ due to the TF coil hoop stress, which produced a large unbalanced radial force at the CS-TF boundary. When $I_\text{CS} \neq 0$, this force is opposed by the hoop force in the CS. At $I_\text{CS} = 0$, model predicted a maximum (hoop) stress component of 920 MPa, as shown in Fig. \ref{fig:jc-stress}. This excludes epoxy as a structural material, but the stress remains below the yield stress of Inconel-718 ($\sim$1000 MPa). While 920 MPa is above the 700 MPa limit at which point $J_c$ is diminished\supercite{Barth2015}, the peak field on coil declines simultaneously and the margin between J$\rm_{op}$ and J$\rm_c$ increases. While the CS is not expected to quench during the current ramp, time-dependent simulation of a realistic geometry is necessary.

The 8 T peak field on the PF coils permitted a higher $\rm J_{ op} = 175\, A/mm^2$ while retaining more than 30\% margin with respect to $\rm J_{c}$\supercite{Molodyk2021}. With a larger current density, the coils can be made smaller, reducing their cost and easing the neutron shielding requirements.  Details of the PF coils are listed in Table \ref{tab:PF_params}. To ensure their ability to generate the desired magnetic geometry is not affected, a PFs were required to be displaced at most 1 cm. This was readily satisfied as the maximum predicted displacement was less than 1.7 mm.
\begin{table}
\begin{center}
\caption{Poloidal field coil parameters}
\label{tab:PF_params}
\resizebox{\columnwidth}{!}{
\begin{tabular}{lllll}
 \hline
 Parameter & PF1 & PF2 & PF3 & Units\\
 \hline
 \hline
Max Current$\cdot$turns & 7.74 & 7.51 & 5.13 & [MA$\cdot$turns]\\
Number of turns & 86 & 84 & 59 & - \\
Height & 0.327 & 0.29 & 0.14 & m\\
Width  & 0.365 & 0.327 & 0.215 & m\\
R & 3.79 & 5.45 & 6.65 & m\\
Z & $\pm$2.25 & $\pm$2.9 & $\pm$1.25 & m \\
Rotation & 108.7 & 86.8 & 30.5 & degree\\
 \hline
\end{tabular}}
\end{center}
\end{table}

\section{Nuclear Analysis}
\label{sec:neutronics}
MANTA's blanket system consists of components to extract energy, produce tritium, and shield other systems from the neutrons generated in the core. This section describes the blanket geometry, material selection, blanket power multiplication factor, volumetric heat deposition, magnet lifetimes limited by high energy neutron fluence, and damage to the vacuum vessel. Limits on the tritium breeding ratio derived from a fuel cycle analysis will also be discussed. 

The design of the blanket system was driven by four key goals in accordance with the NASEM requirements:
\begin{enumerate}
    \item Achieve a toroidal field (TF) coil lifetime of at least 1000 megawatt-years and maximize poloidal field (PF) coil lifetimes to increase economic viability
    \item Maximize the fraction of energy deposited in the blanket to increase reactor efficiency.
    \item Minimize activation of reactor components to reduce health risks and facilitate maintenance;
    \item Reach a tritium breeding ratio (TBR) of at least 1.02 (based off modeling discussed in Section \ref{ssec:neutronics/fuelcycle}).
    
\end{enumerate}

\subsection{Component Geometry and Materials}

A liquid immersion blanket, consisting of molten 2LiF-BeF$_2$ (FLiBe) flowing down and around the vacuum vessel in a toroidally continuous tank, was selected due to the improved reactor serviceability and enhanced TBR relative to traditional blanket designs that rely on tritium breeding modules inside the vacuum vessel containing significant amounts of non-breeding structural material \supercite{Sorbom2012}.

\begin{figure}
    \centering
    \includegraphics[width=0.49\textwidth]{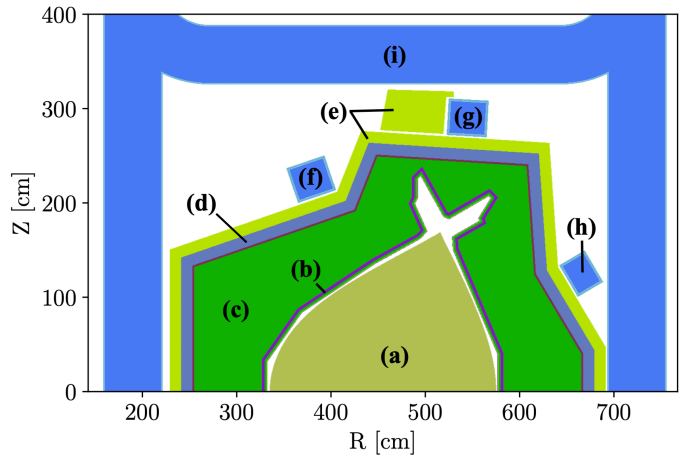}
    \caption{Upper half of a poloidal cross section of the OpenMC model of MANTA. The plasma (a) is surrounded by the vacuum vessel (b). The FLiBe blanket tank (c) is surrounded by layers of WC (d) and B$_4$C shielding (e). PF coils 1-3 are labeled (f-h) respectively. All components are located within the TF coil cage (i). In order to depict all components in greater detail, the upper portion of the TF joints is cut off.}
    \label{fig:openmccs}
\end{figure}

The Monte Carlo code OpenMC\supercite{Romano2015} was employed to model the neutron and photon transport throughout the reactor geometry. These simulations were used to guide the choice of material and geometry for the breeding blanket and other shielding components, as well as the location of the magnets. The model employed a fixed neutron source with neutron production profiles based on the transport modeling of MANTA's fusion core plasma described in Section \ref{sec:STEP}. The global variance reduction methodology MAGIC was used to generate weight windows for converging quantities of interest at the magnet locations. The upper half of a poloidal cross section of the model is shown in Fig. \ref{fig:openmccs}. Aside from the TF coil (see Fig \ref{fig:tf_cross} at a diagram of the TF coil), the geometry is up-down symmetric. 

Two 12-cm-thick layers of shielding material surround the blanket tank, with the inner layer made of tungsten carbide (WC) and the outer layer made of boron carbide (B$_4$C). An additional shield of 42 cm-thick B$_4$C is located outside these layers above the inner divertor leg to provide further shielding of the TF magnets from neutrons streaming through the divertor. A vanadium-chromium-titanium alloy (V-4Cr-4Ti) was chosen for the vaccum vessel (VV) material, which is predicted to be compatible with the blanket if MoF$_6$ is dissolved in the FLiBe to generate a self-healing molybdenum barrier at the vessel-blanket interface \supercite{Sze1986, Muroga2005}. Other candidate vessel materials compatible with the FLiBe operating temperature and activation requirements include oxide dispersion strengthened (ODS) ferritic steels and silicon carbide ceramic composites (SiC-SiC). Given MANTA's modularity, these materials could be explored later in MANTA's life cycle based on material technological readiness levels \supercite{Zinkle2009}.

\begin{figure}
    \centering
    \includegraphics[width=0.49\textwidth]{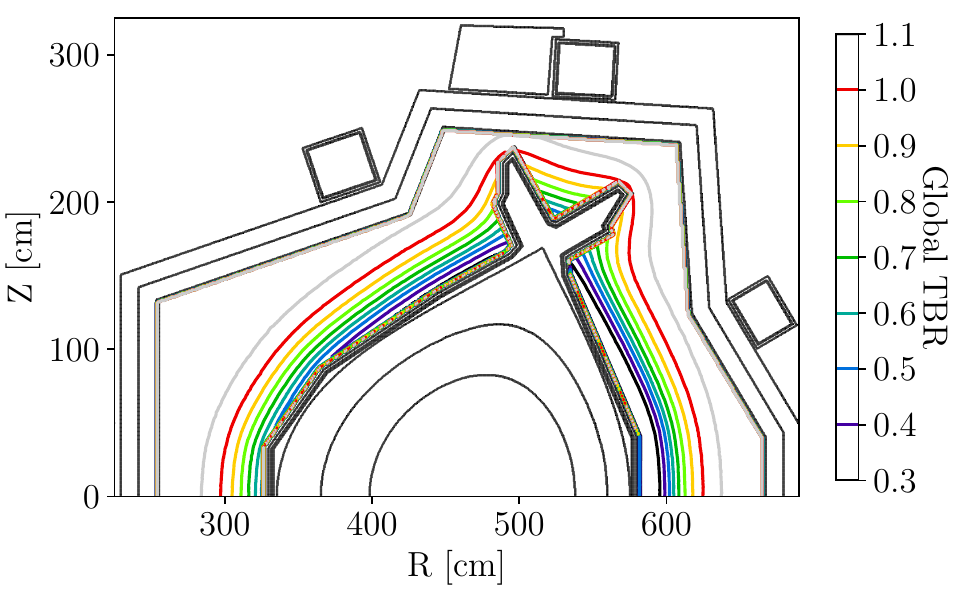}
    \caption{Upper half of a poloidal cross section of MANTA. Overlaid are contours of blanket shapes that achieve a given TBR using the smallest blanket volume possible.}
    \label{fig:tbrcontour}
\end{figure}

By approximately conforming the blanket tank cross section to contours in local heat deposition and TBR, a design was identified that successfully achieves desired magnet shielding and tritium breeding while making the best use of the space available within the TF coils. Fig. \ref{fig:tbrcontour} depicts a skeleton of the vessel components within the toroidal field coils overlaid on a contour plot of the global TBR. The contours indicate the smallest blanket shape possible that would achieve the corresponding TBR. This optimization yielded a TBR of 1.15 and a power multiplication factor (as defined in \cite{Meier1984} as $M_f$) of 1.11, based on heat deposition and lithium reactions occurring in the FLiBe blanket. A summary of the components along with their materials, densities, radial thicknesses, volumes, and volumetric heating is shown in table \ref{tab:neutronicsGeom}. Heating rates assume energy from secondary electrons is deposited locally. The volumes were calculated stochastically in OpenMC.

\begin{table*}[h!]
\begin{minipage}{\textwidth} 
\begin{center}
\caption{Neutronics system components along with their materials, densities, radial thicknesses, volumes, and volumetric heating.}
\label{tab:neutronicsGeom}
\begin{tabular}{ llllll }
 \hline
     Component & Material & Density & Thickness & Volume & Heating \\
      &  & [g/cm$^3$] & [cm] & [cm$^3$] & [W/cm$^3$/MW$_\text{fus}$] \\ \hline \hline
     First wall & Tungsten & 19.3 & 0.3 & $(6.8\pm0.2)\times10^4$ & $(4.6\pm0.3)\times10^{-2}$ \\
     Vacuum vessel & V-4Cr-4Ti & 6.05 & 1.0 & $(2.21\pm0.03)\times10^5$ & $(1.75\pm0.02)\times10^{-2}$ \\
     Cooling channels & FLiBe & 1.94 & 2.0 & $(4.55\pm0.05)\times10^5$ & $(1.18\pm0.02)\times10^{-2}$ \\ 
     Blanket & FLiBe & 1.94 & 20-100 & $(1.367\pm0.003)\times10^7$ & $(2.16\pm0.01)\times10^{-3}$ \\ 
     WC shield & WC & 15.63 & 12 & $(3.25\pm0.08)\times10^6$ & $(9.6\pm0.1)\times10^{-5}$ \\ 
     B$_4$C shield & B$_4$C & 2.52 & 12 & $(3.35\pm0.08)\times10^6$ & $(7.0\pm0.1)\times10^{-6}$ \\ \hline
\end{tabular}
\end{center}
\end{minipage}
\end{table*}

\subsection{Magnet Lifetimes}

The blanket and shielding designs were optimized to maximize the TF and PF coil lifetimes, leading to a blanket shape larger than the contours depicted in Fig. \ref{fig:tbrcontour}. Existing studies have predicted the lifetime high energy (above 100 keV) neutron fluence tolerable by REBCO magnets to be $3\times10^{22}$ neutrons/m$^{2}$ \supercite{Fischer2018, Prokopec2015}. Because the neutron fluence scales linearly with the fusion power of the core (and to account for potential variable power output from MANTA), magnet lifetimes are reported in units of megawatt-years. Dividing by the device fusion power therefore yields the magnet lifetime at that fusion power. Table \ref{tab:maglifetimes} summarizes both the lifetime averaged over the cross section as well as the minimum lifetime of each magnet.

The target lifetime of 1000 MW-yr was far surpassed for the TF coils, a major benefit given their high cost (see Section \ref{sec:econ}). The TF minimum lifetime of ${\sim}3100$ MW-yr corresponds to nearly 7 years of continuous full-power operation at $P_\text{fus} = 450$ MW. Two of the three PF pairs also surpassed 1000 MW-yr, but PF2 is slightly below due to its proximity to the plasma and thus higher neutron fluence. MANTA's environmental cycle is therefore set by PF2, which will require replacement every ${\sim}2$ full-power years. Replacement of individual magnets as necessary can be accomplished quickly relative to other reactor designs due to MANTA's simplified maintenance scheme (Sec. \ref{ssec:magnets/maintenance}).  In Section \ref{sec:econ}, these lifetimes are shown to be sufficient for MANTA to meet its economic targets. 

\begin{table}[h]
\begin{center}
\caption{Mean and minimum lifetimes (in units of MW-yr) for poloidal field (PF) and TF coils based on high energy neutron fluence.}
\label{tab:maglifetimes}
\begin{tabular}{ lll }
 \hline
    Coil & Mean Lifetime & Min. Lifetime \\ \hline \hline
     PF 1 & 3300 $\pm$ 100 & 1320 $\pm$ 80 \\ 
     PF 2 & 3200 $\pm$ 90 & 890 $\pm$ 40 \\ 
     PF 3 & 4900 $\pm$ 150 & 2000 $\pm$ 220 \\ 
     TF & 30400 $\pm$ 400 & 3100 $\pm$ 400 \\ \hline
\end{tabular}
\end{center}
\end{table}

\subsection{Vacuum Vessel Activation and Radiation Damage} \label{subsec:component_activation}

\begin{figure}
    \centering
    \includegraphics[scale=0.5]{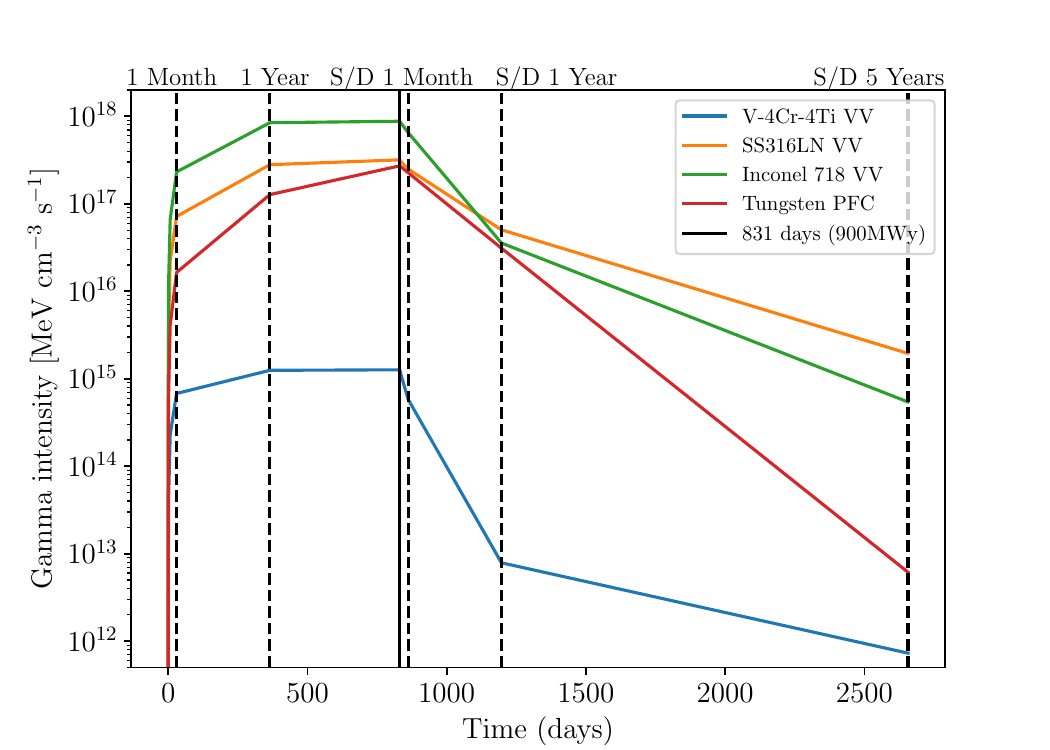}
    \caption{Gamma intensity of the vacuum vessel and plasma facing wall under an irradiation cycle of 900MWy at continuous fusion power of 395MW, followed by 5 years of decay.}
    \label{fig:900MWySpec}
\end{figure}

To estimate the radiological hazard posed by the VV after removal from MANTA, a depletion study was carried out. First, the model was run with the OpenMC transport-coupled operator using tabulated depletion chain data\supercite{ENDFB7} in conjunction with a first-order predictor integrator. The activation analysis used a simplified irradiation schedule, assuming uninterrupted operation at 395MW for 832 days (900MWy). From that point, the integrator was run with no neutron source, allowing materials to decay for 5 years. Photon energy spectra were calculated using material compositions resulting from the activation analysis. The gamma intensity from the vacuum vessel and plasma facing wall are shown in figure \ref{fig:900MWySpec}. V-4Cr-4Ti activation is compared against SS316LN and Inconel 718, two other commonly proposed VV materials. Activation of the V-4Cr-4Ti VV is at least 3 orders of magnitude lower than the other candidate materials at all times, reducing the radiological hazard during maintenance and disposal.

The average displacements per atom (DPA) in the vacuum vessel was determined to be 2.24 per 100 MW-yr using the model proposed by Norgett, Robinson, and Torrens \supercite{norgett1975}. The maximum local DPA rate in the VV was found to be 2.91 per 100 MW-yr. Assuming replacement every 2 full-power years, the resulting average DPA of 19.7 and maximum DPA of 20.5 is expected to be tolerable by the V-4Cr-4Ti\supercite{SMITH1995399, SMITH2000213,Zinkle2009}. However, this must be verified in situ or with conditions more representative of the first wall.

\subsection{Fuel Cycle Analysis}
\label{ssec:neutronics/fuelcycle}
MANTA's fuel cycle was modeled in the MATLAB Simulink toolbox via an open-source fuel cycle analysis code\supercite{samuele_meschini_repo}. This model accounts for (among other factors) tritium startup inventory, fuel burn fraction in the plasma, and tritium losses in the system. Unlike other tokamak systems, a fully functioning tritium fuel cycle has yet to be developed or tested, so the following model employs conservative estimates for model parameters (e.g. tritium losses, system inefficiency, startup inventory, and availability factor).

The fuel system consists of two largely independent cycles: the inner cycle (consisting of the wall, blanket, divertor, etc., responsible for tritium breeding) and the outer cycle (consisting of fuel separation, cleanup, and storage). This design, as well as the estimates for pumping speeds and inherent system losses, were adapted from \cite{Meschini2023}.

Two of the most sensitive parameters of the fuel cycle are the startup and reserve tritium inventories. While the startup inventory can be calculated from other variables, the reserve inventory (i.e. how much extra tritium is stored at the plant) necessitates a balance between radiation safety, cost, and the likelihood of a system failure. Given these constraints, a reserve inventory of 75g was set, resulting in a required startup inventory of 440g was calculated. As tritium fuel systems further mature, these numbers will likely drop, relaxing both economic and safety concerns.

\begin{figure}
    \centering
    \includegraphics[width=1\columnwidth]{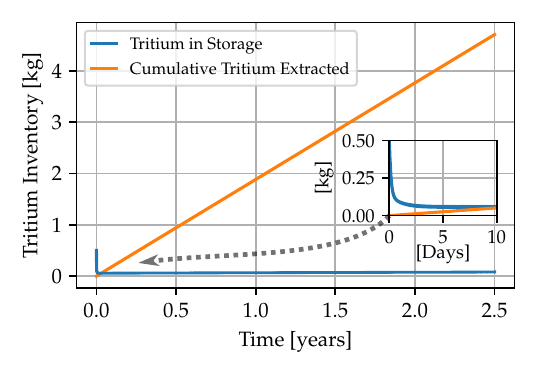}
    \caption{Tritium inventory and production over the plant's lifetime, assuming a conservative TBR of 1.10. Quickly after beginning operation, the inventory equilibrates to the 75g reserve inventory. Assuming a constant extraction rate, more than 4 kg of tritium are produced in 2.5 years}
    \label{fig:trit_extr}
\end{figure}

The absolute minimum TBR necessary to sustain plant self-sufficiency with a 440g startup inventory was found to be 1.02. The predicted TBR of 1.15 therefore allows excess tritium production for use to start up future plants. To provide more conservative results, the fuel-cycle simulations assumed a TBR of 1.10. This accounts for unexpected system inefficiencies as well as the future addition of systems that will need to cut through the blanket, thereby reducing its volume and the system's TBR. This estimate is justified in part by OpenMC simulations that have shown a conservative model of a feedthrough for the RF heating system reduces TBR by about 1\%. MANTA's fuel system will equilibrate within a week (allowing quick return to normal operation following a maintenance cycle) and generate enough tritium to start up another MANTA-class device every six months (see Fig. \ref{fig:trit_extr}). 

\section{Plant Operation}
\label{sec:balance_of_plant}

To ensure that MANTA met the NASEM requirement for minimum electricity production (${\geq 50}$ MWe over {$10^4$ s}) and electricity gain ($Q_e \geq 1$), an initial balance of plant and thermal cycle scoping was completed. Multiple thermal cycles were considered, including Brayton, sub-critical Rankine, and super-critical Rankine. 

The final design used a steam Rankine cycle fueled by a two-stage molten-salt loop with a thermal storage system to provide constant thermal power to the power cycle for constant electricity production. 
The plant layout and molten salt storage system are detailed in section \ref{sec:bop_layout_mss}. A steady state net electricity production of at least 90 MWe was achieved at an electrical gain of $Q_e = 2.4$ for both sub- and super-critical Rankine cycles. The final cycle selection and performance parameters are discussed in Section \ref{sec:bop_power_cycle}.

\subsection{Input Constraints and Performance Parameters}
\label{sec:bop_constraints}
Aside from the performance requirements set by the NASEM report, the largest driver of MANTA's thermal plant design was the input thermal power generated by fusion, $P_\text{fus} = 450$ MWth, and the additional heat produced by exothermic reactions in the blanket. A finite recirculating power was also required for the auxiliary RF heating ($P_\text{aux,e} = 57$ MWe) and cryogenics ($P_\text{cryo,e} = 1$ MWe) subsystems. To generate a steady state power output, the molten-salt storage system must store enough energy during the $15$ minute pulse to provide sufficient power to the turbine during the ${\sim}2$ minute interpulse time.  Finally, the maximum operating temperature of the vacuum vessel material, V-4Cr-4Ti, under radiation and thermal load set an upper limit of 600\degree C on the temperature of the hot (FLiBe) leg of the thermal cycle \supercite{Zinkle2009}.

The total net electrical power, $P_\text{e,net}$, was calculated for a Brayton, sub-critical Rankine, and super-critical Rankine power cycle from 
\begin{equation}
    P_\text{e,net} = P_\text{turb,e} - P_\text{systems,e},
    \label{eq:pe}
\end{equation}
where $P_\text{turb,e}$ is the electrical power generated by the power cycle turbine and $P_\text{systems,e} = P_\text{aux,e} + P_\text{cryo,e} + P_{\rm pumps,e}$ is the electrical power to run the subsystems, namely the auxiliary heating, cryo, and pumps/compressors. The turbine mechanical-electrical power conversion efficiency was assumed to be 1, and the pumps' electrical-mechanical power conversion efficiency was assumed to 0.75. The electricity gain factor $Q_e$ is defined such that $Q_e > 1$ corresponds to net electricity gain, matching the definition in the NASEM report\supercite{NASEM_report}.
\begin{equation}
    Q_e = P_{\rm turb,e} /P_{\rm systems,e}.
    \label{eq:qe_general}
\end{equation}
A thermodynamic analysis of the overall MANTA plant design was completed to calculate these quantities.

\subsection{Plant Layout and Molten Salt Storage System}
\label{sec:bop_layout_mss}
MANTA's thermal plant consists of a primary FLiBe loop in which heat is deposited due to fusion, a secondary molten salt loop to isolate the power cycle working fluid from contamination, and a power cycle that generates electricity via a turbine. The thermal storage system was incorporated into the secondary loop to ensure constant thermal power to the power cycle both during a discharge (${\sim}$15 minutes long) and between pulses (${\sim}$2 minutes long).

A constant thermal power has the additional benefits of keeping the turbine at its maximally efficient, rated power output and keeping the pressures and temperatures in the system close to steady-state. The secondary loop and the storage system employ a molten salt mixture of 60\% NaNO$_3$/40\% KNO$_3$, which is a typically used for thermal storage in solar energy plants. This salt was selected here for its boiling and recrystallization points being above the hot-leg temperature of the primary-secondary loop molten salt heat exchanger and below the cold-leg temperature of the secondary-power cycle heat exchanger, respectively \supercite{inl_liquid_salt}. 

\begin{figure}[h]
    \centering
    \includegraphics[width=\columnwidth]{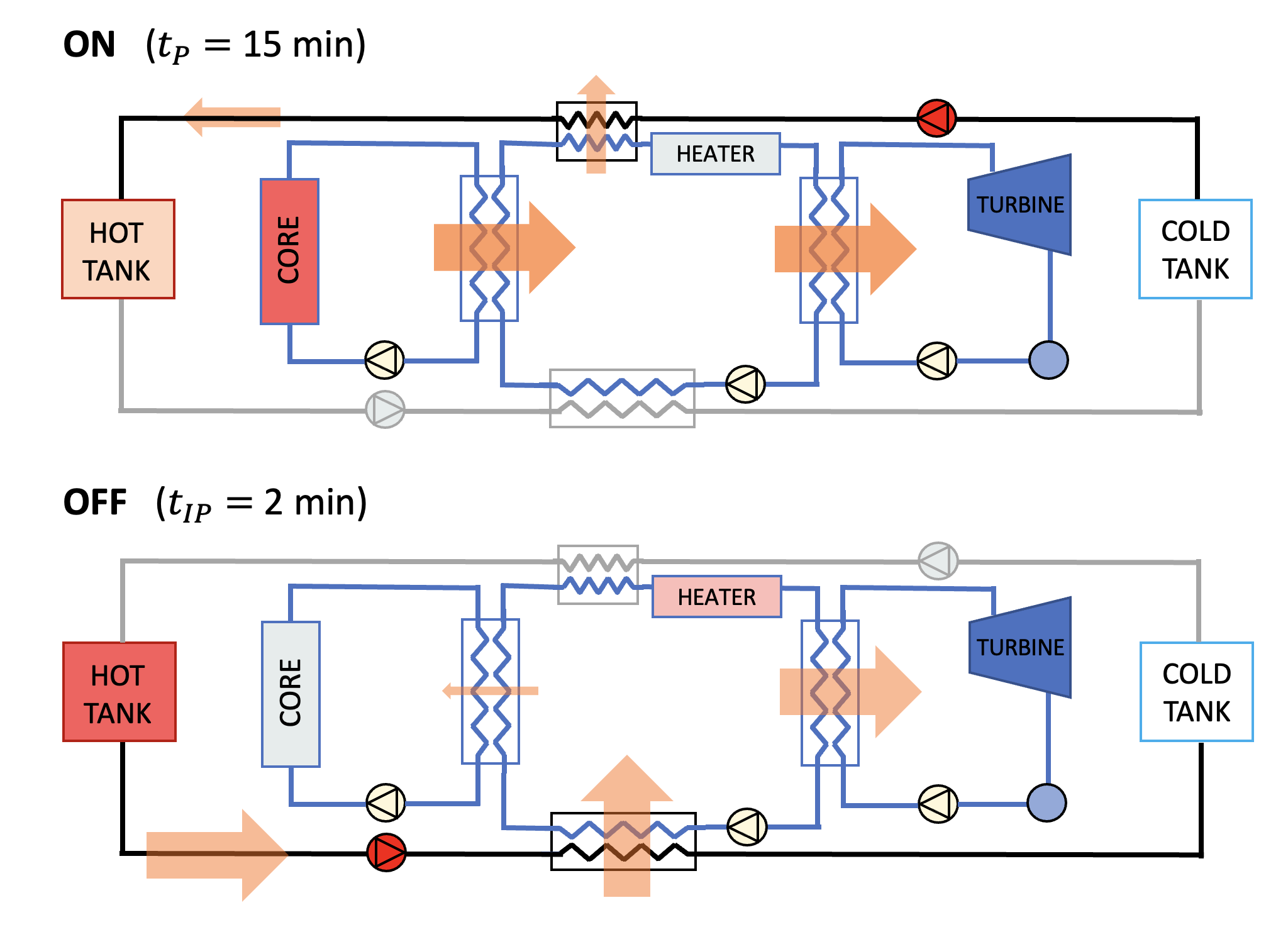}
    \caption{Molten salt loops and power cycle diagram for pulse and interpulse operation. Cycles are connected through heat exchangers between them and to the surrounding storage system. Circumscribed triangles represent pumps: yellow pumps are continuously run, and red/gray pumps are on/off depending on the phase of operation.}
    \label{fig:molten_salt_storage}
\end{figure}

A diagram of the complete thermal plant, including the storage system is shown in Fig. \ref{fig:molten_salt_storage}. The flow of power is highlighted with orange arrows. During a discharge (upper plot), thermal power is captured or generated by the primary cycle and transferred into the secondary cycle. A fraction of this power is tapped from the secondary salt loop and stored in the hot tank via exchange of heat with the cold tank, reducing the thermal power to the power cycle. Between pulses (lower plot), heat is taken from the hot tank to maintain a constant thermal power to the power cycle. In order to maintain the molten salts at constant temperatures between pulses, a small fraction of power is transferred to the FLiBe cycle to balance thermal losses, and a heater is run with some of the power allocated to the auxiliary plasma heating systems, which is not needed during interpulse operation.

To calculate the pulse-averaged thermal power sent to the power cycle, the flow of heat through the various sub-cycles was considered. The gross thermal power from the blanket, including the multiplication factor $M_f$ calculated by the OpenMC neutronics simulations is
\begin{equation}
P_{\rm th,gross} = \left[ \frac{E_{\alpha}}{E_{DT}} + M_f\frac{E_n}{E_{DT}} \right] P_\text{fus} + P_\text{aux} , 
\end{equation}
where $E_{\alpha} = 3.5$ MeV, $E_n = 14.1$ MeV, and $E_{DT} = 17.6$ MeV are the energies of fusion-born $\alpha$-particles, fusion-born neutrons, and total released by D-T fusion. $P_\text{aux} = 40$ MW is the RF power after tetrode electricity-to-RF efficiency. The pulse-averaged thermal power is then
\begin{equation}
\resizebox{.85\columnwidth}{!}{$
    \langle P_{\rm th} \rangle_{\rm pulse} = \eta_{HX}^2 P_{\rm th,gross} \left [ 1- \frac{1}{1+\eta_{HX}^2 t_P/t_{IP}}\right],$}
\end{equation}
where the heat exchanger efficiencies $\eta_{HX}$ were assumed to be 0.95. The first term represents the thermal power without the storage system contribution, which passes through the primary-secondary loop and secondary-power cycle heat exchangers. The second term is the power tapped stored in the thermal storage system during pulses. The two additional passes through heat exchangers when the core is off result in a temperature drop, which is recovered with the heater. 

\subsection{Power cycle choice}
\label{sec:bop_power_cycle}

Three thermal cycles were considered for MANTA: Brayton, sub-critical Rankine, and super-critical Rankine. Brayton cycles typically operate using air or helium as a working fluid, with helium being preferred for its high heat capacity and chemical inertness \supercite{mitsubishi_design_1996,no_review_2007}. Although Brayton cycles are currently used in natural gas power plants and have been proposed for use in high temperature gas cooled nuclear power plants, their effectiveness relies on achieving a high temperature differential across the cycle. Modern Rankine cycles are often supercritical and operate at higher inlet temperatures and pressures to increase efficiency over sub-critical cycles \supercite{pwr_plant_eng}. There are also advantages to using a single-phase, super-critical working fluid, including decreased heat exchanger complexity and reduced degradation of critical turbine components due to water vapor. 

A thermodynamic analysis of basic Brayton and Rankine cycles was performed to evaluate performance. The Brayton cycle is closed, has one turbine stage, and uses regeneration to capture heat leaving the turbine, with 80\% efficacy \supercite{eng_thermo_8}. The base pressure is 10 bar, which is relatively high but typical of helium Brayton cycles \supercite{no_review_2007}. At the MANTA design point, the optimized compression ratio was 2.0. The sub- and super-critical Rankine cycles use two turbine stages and include regeneration via an open feedwater heater, which taps some fraction of the heated working fluid from the high-pressure turbine outlet and mixes it with the outlet of the first pump coming after the condenser. The thermodynamic states and regeneration fraction are optimized for maximum thermal efficiency. The maximum pressure is set to 150 and 300 bar for the sub- and super-critical cycles, respectively. The maximum temperature for all cycles is set to {560\degree C}, which considers a {20\degree C} temperature drop across the FLiBe to solar salt and the solar salt to power cycle heat exchangers. This is a conservative temperature drop given the low technological readiness level of molten salt heat exchangers, and is expected to be reduced as these components are further developed. All thermodynamic state analysis and optimization was performed in Python using the \texttt{pyXSteam}\supercite{pyxsteam} and \texttt{PYroMat}\supercite{PYroMat} libraries. The details of the power cycle models, including thermodynamic properties at the final design point are provided in \ref{appx:thermo}. 

\begin{figure}
    \centering
    \includegraphics[width=0.5\textwidth]{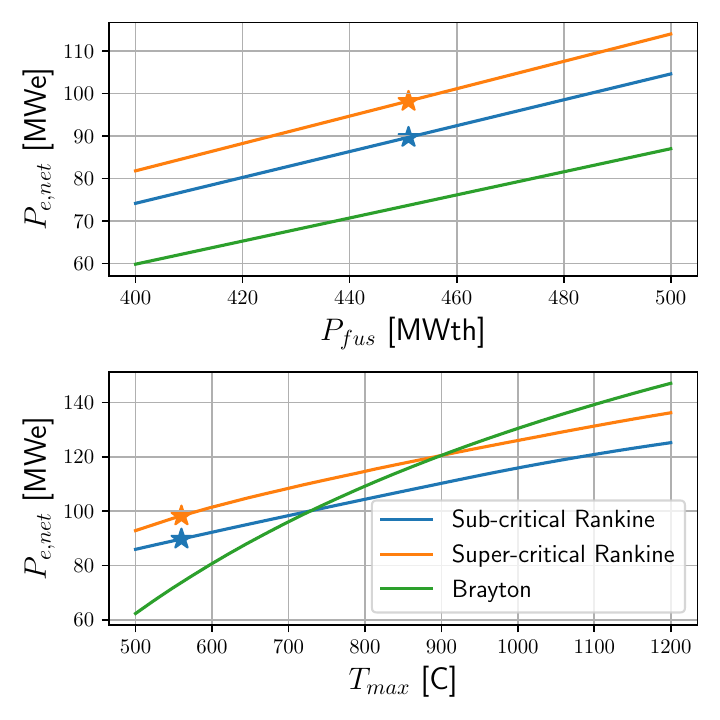}
    \caption{Upper: $P_\text{e,net}$ for 3 different power cycles over a range of $P_\text{fus}$ at fixed $T_\text{max}$=560 \degree C. Lower: $P_\text{e,net}$ for 3 different power cycles over a range of $T_\text{max}$ at fixed $P_\text{fus}$= 450 MWth. The starred points indicate the final design points for the selected power cycles.}
    \label{fig:cycle_comparison}
\end{figure}

Fig. \ref{fig:cycle_comparison} shows $P_\text{e,net}$ for each of the cycles over scans of $P_\text{fus}$ (upper plot) and the power cycle maximum temperature, $T_\text{max}$ (lower plot). At each value of $T_\text{max}$, the Rankine cycle regeneration parameters and Brayton cycle compression ratio are re-optimized for consistency. The $T_\text{max}$ scan shows that at the operating $T_\text{max}$ of the power loop, both Rankine cycles outperform the Brayton cycle, though Brayton cycles can provide a higher power out at higher temperatures. The $P_\text{fus}$ scan at fixed $T_\text{max}=560$\degree C shows both Rankine cycles outperforming a Brayton cycle and that for a wide range of fusion powers around the 450 MWth design point, $P_\text{e,net}$ scales nearly linearly with $P_\text{fus}$. These scans resulted in the choice of a Rankine cycle for MANTA. 

And the large gains in $P_\text{e,net}$ at higher $T_\text{max}$ demonstrate the benefits of a VV capable of withstanding such temperatures. Alternative VV materials such as oxide dispersion strengthened (ODS) ferritic steels or silicon carbide ceramic composites (SiC-SiC) could be explored later in MANTA's life to increase electricity production as the technological readiness levels of these materials increase.

Important performance parameters for each cycle are reported in Table \ref{tab:thermo_cycle_output}. The thermal efficiency is calculated for each cycle as the net thermal power out divided by the thermal power in
\[
\eta_{th} = (P_{\rm turb,th} - P_{\rm pump/comp,th})/\langle P_{th} \rangle_{\rm pulse}.
\]
This definition does not reflect the benefit of coupling the mechanical shaft of the turbine and the compressor in the Brayton cycle, as compared with running electric pumps with a certain efficiency in the Rankine cycles, but this is considered when calculating $P_\text{e,net}$. The super-critical Rankine cycle has the best performance, with the highest $\eta_{th}$, $P_\text{e,net}$, and $Q_e$. The Brayton cycle has a higher $P_{\rm systems,e}$ than either Rankine cycle due to the increased power cost of the gas compressor compared to the water pumps. Notably, all of the considered cycles exceed the NASEM report requirements while conforming to the constraints imposed by MANTA's various subsystems. Given their superior performance, both Rankine cycles are found to be suitable for MANTA's electricity generation. The sub-critical cycle was ultimately chosen for the possibility of brownfield siting, which is further discussed in Sec.~\ref{sec:econ}.
\begin{table}[]
    \centering
    \resizebox{\columnwidth}{!}{
    \begin{tabular}{ lllll }
 \hline
  & \makecell{Sub-critical \\ Rankine} & \makecell{Super-critical \\ Rankine} & Brayton & Units \\
 \hline
 \hline
  $\eta_{th}$ & 0.36 & 0.39 & 0.33 & - \\ 
  $P_\text{systems,e}$ & 63 & 69 & 257 &MWe \\ 
  $P_e$ & 90 & 98 & 74 & MWe \\
  $Q_e$ & 2.4 & 2.4 & 1.28 & MW$_\text{th}$ \\
 \hline
    \end{tabular}
    }
    \caption{Performance parameters at the operating point for sub-critical Rankine, supercritical Rankine and Brayton cycles.}
    \label{tab:thermo_cycle_output}
\end{table}

\section{Economic Analysis}
\label{sec:econ}
The NASEM report states that a viable pilot plant must achieve an overnight cost of less than US\$5 billion, writing ``If the private sector, even with government backing ..., will not accept a total price past US\$5 billion to US\$6 billion for generating technology already demonstrated at scale, then it will certainly not accept this for the pilot plant.''\supercite{NASEM_report}.
To ensure that MANTA met this requirement, a techno-economic analysis was performed. Critically, a pilot plant must also provide confidence that an $N^\mathrm{th}$-of-a-Kind, commercial power plant of a similar design has a path to profitability. With this goal in mind, the levelized cost of electricity (LCOE) for MANTA and a scaled-up power plant version was calculated, accounting for costs across the project lifetime as well as revenue streams.

\subsection{Overnight Cost Assessment}
Leveraging a bottom-up techno-economic model, an overnight cost of US\$3.4B was calculated, more than meeting the NASEM report requirement. As will be detailed below, this value represents the sum of tokamak costs (Table \ref{tab:componentCosts}), other direct costs (physical infrastructure, Table \ref{tab:Direct costs}), and indirect costs (service costs, Table \ref{tab:Indirect Costs}). A contingency of 10\% was added to the total budget, reflecting standard industry practice of building in cost ``head room'' to increase confidence in a large-scale capital budget\supercite{FRA_Capital_2016}. The US\$3.4B overnight cost translates to a unit cost of $\simeq$US\$38 million / MWe. While MANTA was not designed as a commercial power plant, this is within a factor of two of the Dominion Energy Coastal Virginia Offshore Wind 12~MWe (pilot) plant project built in 2020 \supercite{woodruff_conceptual_2017,NASEM_report}.

\begin{figure}[h]
\centering
\includegraphics[width=1\columnwidth]{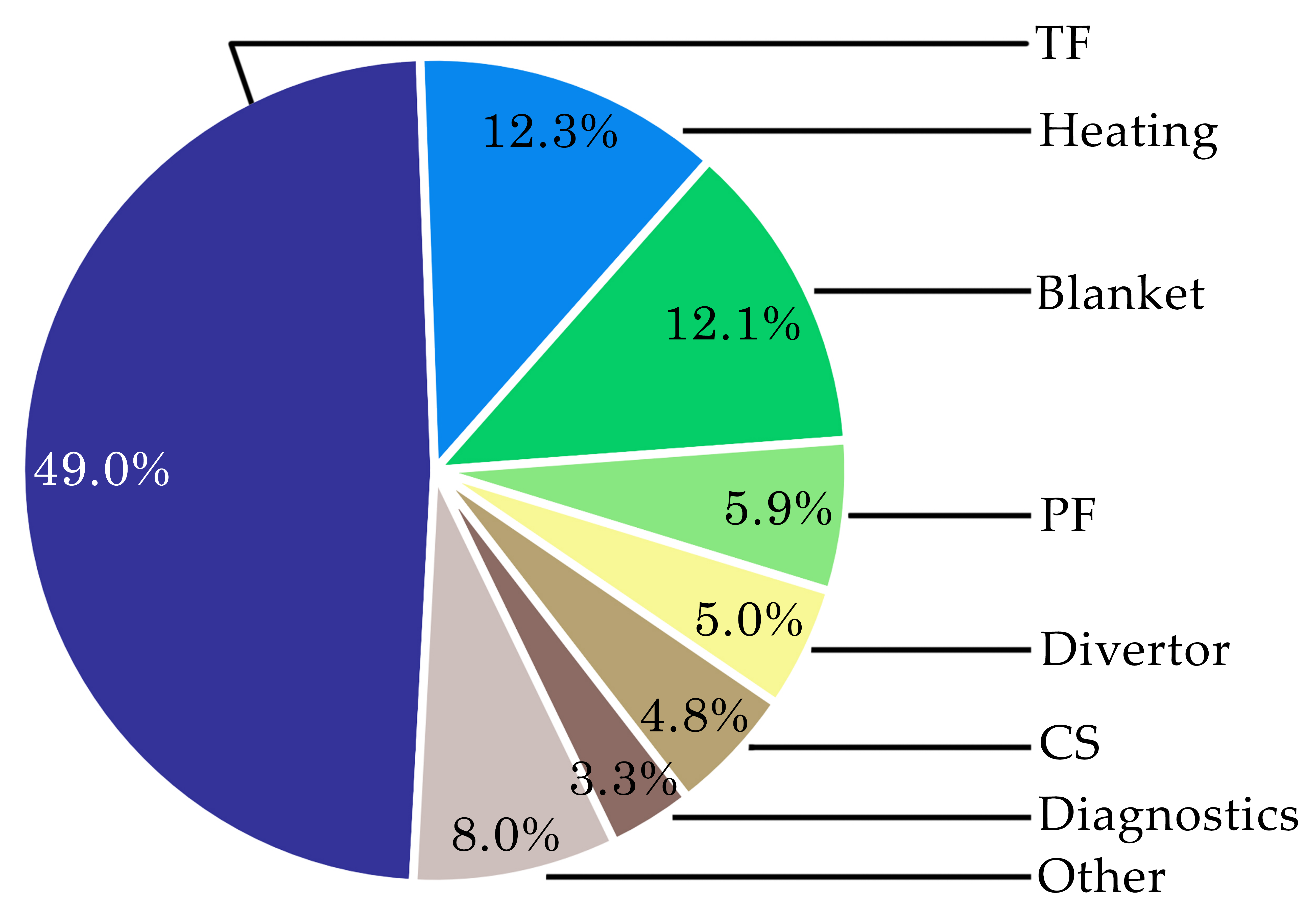}
\caption{Break down of the major systems contributing to the cost of the tokamak. The tokamak makes up 89\% of MANTA's total cost.}
\label{fig:Overnight_cost_pie}
\end{figure}

The tokamak itself was predicted to cost US\$3.1B, ${\sim}89\%$ of the total overnight cost. The largest cost driver is the toroidal field (TF) coils at US\$1.5B, as illustrated by Fig. \ref{fig:Overnight_cost_pie}. To cost the tokamak, subsystems were broken into their constituent components wherever possible. Using simplified geometric models of each component, the component mass was then estimated. Total material costs were calculated according to Table \ref{tab:materials costs}. To best ensure that the overnight cost stayed within NASEM limits, a conservative model was applied for converting material cost to fabricated component cost. Labor costs (L) were estimated as proportional to the material costs (M), with a fabrication factor of 3x for traditional components and 5x for superconducting magnets:
\begin{equation}
    M + L = \text{fabrication factor} * M
\end{equation}
This approach is similar to the cost per tonne scaling laws used in \cite{Sorbom2012} and \cite{meade_comparison_2002}. However, it allowed the impact of materials selection, and their dramatic range in costs, to be explored. Further assessing the techniques necessary for manufacturing each component and using these to refine the labor cost estimates would be a valuable area of future research.
The turbine plant cost was estimated from \cite{INL_HTGR_costing}, and remaining direct and indirect costs were taken from the ARPA-E report on fusion power plant costs \supercite{woodruff_conceptual_2017}.

\begin{figure}
    \centering
    \includegraphics[width=\columnwidth]{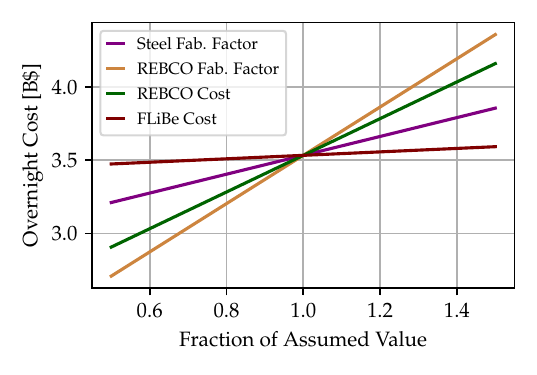}
    \caption{Sensitivity analysis ($\pm 50\%$) of key values assumed in techno-economic modeling. Within this range of values, costs remain under the targeted maximum of US\$5B.}
    \label{fig:assumption_scan}
\end{figure}

To gain confidence in this costing estimation, a sensitivity study scanning $\pm 50\%$ was performed on the four key cost drivers with most uncertainty: the fabrication factor for traditional components, the fabrication factor for superconducting components, the cost of REBCO, and the cost of FLiBe, the results of which are seen in Figure \ref{fig:assumption_scan}. Throughout this range of values, the overnight cost remained less than the US\$5B, increasing the confidence that MANTA meets this requirement. 

It is assumed for this analysis that MANTA is built at a ``brownfield'' land site (that of an unused fossil fuel plant).  This option saves on costs related to the sub-critical Rankine cycle, the electrical plant, and around half of the building construction costs. While legacy plant decommissioning is an additional cost, approximately US\$120k/MWe for a coal plant \supercite{coal_decomissioning}, brownfielding saves an estimated US\$400 M, or around  5\% of the total overnight cost vs a ``greenfield'' (empty land) site, although this could be a more generalizeable use case.

It is also important to recognize the non-monetizable benefit to brownfielding in the energy justice implications of retaining jobs within communities dependent on fossil fuel plants and putting often marginalized communities at the forefront of cutting edge scientific development. 
Furthermore, the relatively small contribution to the capital and operations budget of construction and personnel (see Tables \ref{tab:Indirect Costs} and \ref{tab:Operational_Costs} in the Appendix) and extra room in the overall budget indicate that the jobs created to build and run a MANTA-class reactor could provide good pay and benefits (e.g. ``Good paying, union jobs''\supercite{Biden_2021} meeting prevailing-wage rates).
This is a further component of a just transition to a clean energy economy\supercite{Biden_2021_2} and one with positive implications in both finance (maintaining eligibility for electricity production tax credits\supercite{IRA_2022}) and policy (maintaining public support for fusion energy\supercite{Hoedl_2022}).

\subsection{Beyond NASEM: MANTA Provides Operational Certainty}

While MANTA was not designed to be or directly extrapolate to a commercial powerplant, the design and techno-economic model can help further identify a path to commercially viable fusion energy.
Beyond fulfilling the NASEM requirements, this involved an accounting of MANTA's lifetime costs and revenue streams. The scalings of these factors are succinctly captured by MANTA's levelized cost of electricity (LCOE), the minimum wholesale price of electricity at which the project becomes profitable:
\[
LCOE = \frac{\sum_t \frac{C_t}{(1+d)^t}}{\sum_t \frac{P_{e,net}}{(1+d)^t}},
\]
where $C_t$ are the costs at a specific point in time, $P_{e,net}$ is the net electricity generated at that time, and $d$ is the discounting factor (an industry standard 7\%~\supercite{EIA_Discounting}, necessary to capture the opportunity cost of an investment). The costs included are: financing, personnel, yearly capital degradation, magnet and vacuum vessel replacement, and fueling/selling tritium. See 
Table \ref{tab:Operational_Costs} for further details, and Section \ref{sec:balance_of_plant} for the derivation of $P_{e,net}$. 
A conservative ``learning'' function of 
\begin{equation}
L=0.6+0.4\times^{\left[n^{th}~unit/5\right]}
\end{equation} is assumed, where $n$ is the total number of a given component produced. This slightly reduces the cost of component fabrication over time\supercite{Rubin_21}. No tax credits or any effective corporate tax rate (even when likely applicable, see Table \ref{tab:Operational_Costs}) are assumed as these important factors are too geographically and legislatively dependent. A 3\% inflation rate is assumed.

The gross revenue for an 8.5 year project, including selling both tritium and electricity, is estimated to be US\$205M, and a gross loss of US\$512M (both values include inflation and discounting factors), giving an overall cost of ${\sim}$US\$3.7B. This remains under the upper limit of the overnight cost, although without profitability. The cumulative inflow and outflow over time is shown in Fig \ref{fig:LCOE}. During periods of magnet replacement the electricity production drops to zero (following Fig. \ref{fig:cool_time_diagram}).

\begin{figure}
    \hspace{-3ex}
    \includegraphics[width=1.1\columnwidth]{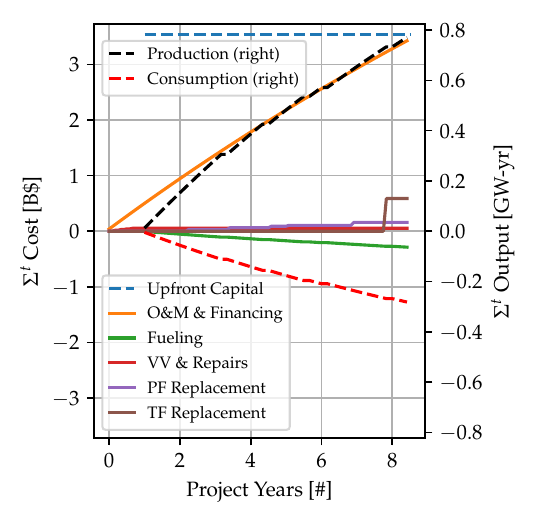}
    \caption{Cumulative cost and electricity inflow/outflows over time for an 8.5 year project, assuming a 1 year construction period. Upfront capital cost shown for reference. Notice the large cost jump at the 1$^\mathrm{st}$ TF replacement timepoint and the zero electricity production during magnet replacement (e.g. year \#3). See table \ref{tab:Operational_Costs} for further details. Tritium sales represented as a negative cost}
    \label{fig:LCOE}
\end{figure}

The LCOE framework gives a simple intuition as to how model parameters affect the overall economic viability, although without self-consistent physics models. For instance, Fig. \ref{fig:LCOE} shows the sale of tritium and replacement of the TF to be key revenue/cost drivers. Scanning over the market price of tritium and the REBCO fabrication labor multiplier cost (similar to Fig \ref{fig:assumption_scan}) show that a 75\% decrease in the former (e.g. due to MANTA's predicted 1.8kg/yr net tritium production saturating the current 2.7kg/yr global tritium production market\supercite{PEARSON20181140}), could be balanced by a 30\% decrease in the latter by improvements in engineering efficiency. Both of these changes generated a factor of $\times$1.6 change in LCOE. 

Finally, the techno-economic model indicated areas of improvement necessary to move from a First-of-a-Kind to a commercially viable $N^\mathrm{th}$-of-a-Kind device. The LCOE model was used to investigate a hypothetical device scaled up in power based off of MANTA. Scaling to a 550 MW core (just within the bounds of the POPCON and below the Greenwald limit, see Fig. \ref{fig:POPCON}) and a 30 year project gives an LCOE of US\$396/MW-hr, assuming current tritium prices. This is significantly too high to be economically competitive, higher even than existing yet novel carbon-neutral technologies, such as offshore wind (\$136.5/MW-hr estimated for projects entering into service in 2027\supercite{EIA_LCOE_Projections}). 

However, the model suggests a direction for a design point of a future MANTA-based reactor which would move closer to profitability: higher power and longer magnet lifetimes
.  If the TF \& PF lifetimes were extended to exceed the lifetime of the 550~MW project, the LCOE drops by 56\%. Similarly, increasing the power cycle efficiency by running the working fluid at 900C (with a vacuum vessel that could handle such a temperature, such as SiC/SiC\supercite{Zinkle2009}) decreases the LCOE an additional 10\%, making this hypothetical plant comparable to offshore wind's price point.  Thirdly, the POPCON model outlined in Section \ref{sec:core} suggests accessing a higher $P_\text{fus}$ regime should be possible without a significant increase in $P_\text{SOL}$ or auxiliary heating power (and accompanying RF heating cost), with a similarly dramatic improvement in LCOE (as $P_\text{fus}$ approaches e.g. $\sim1~\mathrm{GW}$). 

While such operating points have not been self-consistently modeled, it is encouraging that the most detailed economic costing model for a fusion reactor published to date predicted an LCOE even remotely close to profitability. Furthermore, the fact that this is achieved based off extrapolation from a device not optimized for commercial power production and that a path appears to exist towards commercial viability of the high-field NT tokamak concept is highly promising.

\section{Conclusion}

\label{sec:conclusion}

By leveraging negative triangularity and radiative, ELM-free operation to take a ``power-handling first" approach, MANTA (Modular Adjustable Negative Triangularity ARC-class) scales the tokamak concept to a fusion pilot plant (FPP) while maintaining readily survivable heat fluxes on the divertor targets. An extensive integrated modeling workflow confirmed MANTA satisfies the requirements of an FPP for demonstration of the path to commercial viability of nuclear fusion, as detailed in the NASEM report\supercite{NASEM_report}. These criteria are summarized in Table \ref{tab:NASEM_compare}, where MANTA is seen to surpass all metrics. Additionally, MANTA's environmental cycles last ${\sim}$2 full-power-years, making the requirement of operation through several environmental cycles possible in a reasonable amount of time and before the lifetime of the TF coils is reached. 
\begin{table*}
\begin{center}
\caption{Comparison of NASEM criteria and those achieved by MANTA}
\label{tab:NASEM_compare}
\begin{tabular}{lllll}
 \hline
 Parameter & NASEM Requirement & MANTA \\
 \hline
 \hline
 $Q_{E}$ & 1 & 2.4 \\
 $P_{e, net}$ [MWe] & 50 & 90\\
 $TBR$ & 0.9 & 1.15  \\
 Overnight cost (USD) & \$5 Billion &  \$3.4 Billion\\
 \hline
\end{tabular}
\end{center}
\end{table*}

Beyond these criteria, MANTA is designed around its role as a pilot plant, where modifications to both the device itself and the operating point are expected. MANTA's use of demountable TF coils, a liquid immersion FLiBe blanket, and an oversized cryosystem permit relatively rapid replacement of reactor components, ideal for the prototyping of fusion technology. MANTA's fusion power can also be adjusted while maintaining constant $P_\text{SOL}$ through control of the density, allowing for a flexible operating point. Together, these two features significantly enhance MANTA's effectiveness as an FPP.


The most essential area of future work will be continuing NT studies on existing devices. Compared to positive triangularity, negative triangularity is far less understood. While MANTA's success in meeting the NASEM targets together with previous work\supercite{Kikuchi_2014, medvedev2016single,Kikuchi_2019, Medvedev_2015} show the plausibility of NT pilot/power plants, further experimental data, especially with regards to radiative ELM-free plasmas, is required to provide greater confidence that NT can scale to a reactor-class tokamak. 

\section{Contributions}
G.R. led the writing of this work;
D.A., S.B., A.R.D., A.D.M., M.C.C., R.C., L.C., J.J., J.v.d.L, M.A.M., A.S., M.T., A.V., A.M.W., and H.S.W. contributed to the writing of this work;
G.R and A.S. oversaw project management; J.W. designed the SOLIDWORKS CAD models of MANTA;
0D core solution found by A.S. with contributions from S.B., L.C., R.D., J.D.J, P.L., J.v.d.L, and M.P;
CHEASE equilibria generated by R.D., J.v.d.L., and H.S.W.;
Initial RF solution identified by J.D.J and S.J.F and finalized by J.v.d.L. and S.J.F.;
Initial transport scoping completed by S.B., L.C., and M.P.; 
Final transport solution scoped and calculated by H.S.W. with contributions by A.S.;
L.C., J.D.J., and M.A.M worked on core-edge integration;
G.R. generated the FreeGS equilibrium and initial PF coil design; 
N.d.B optimized the PF coils;
D.A. and M.A.M. developed the UEDGE solution with contributions from H.C.; 
Divertor target FLiBe cooling modeled by J.W. with contributions from C.C.;
A.D.M. and A.M.W. developed the maintenance scheme and TF coils;
A.R.D and H.S.W modeled the electromechanical properties of the CS and PF coils;
J.J. and M.T. analyzed magnet lifetimes with contributions from J.L.B.;
J.J. investigated vacuum vessel activation with contributions from J.L.B.; 
Tritium fuel cycle calculations completed by N.D.;
Pulse duration calculated by G.R. with contributions from A.R.D. and S.M.; 
Balance of plant analyzed by M.C.C. and A.V. with contributions from S.M.;
A.S. calculated the overnight capital cost; 
LCOE analysis completed by R.C.; 
S.J.F, C.J.H, N.R.M, and P.R.F mentored the core group; 
A.O.N and M.W mentored the divertor/power-handling group; 
T.M. mentored the magnets group; 
S.F., E.P., and S.S. mentored the neutronics group; 
R.B. mentored the economics and balance of plant groups; 
and C.P.S. and D.G.W. oversaw this work.

\section{Acknowledgements}
The authors are grateful  to all other course members of MIT 22.63/CU-APPH 9143 and for the expert advice of Matt Reinke, Nicolo Riva, Sergey Kuznetsov, Tony Qian, Adam Kuang, Jacob Schwarz, Rui Vieira, Charles Forsberg, Ted Golfinopoulos, as well as to Sean Ballinger for use of his UEDGE post-processing tools. This work was supported in part by US DOE grants DE-FC02-04ER54698, DE-FG02-86ER53222, DE-FG02-91ER54109, DE-SC0007880, DE-SC0014264, DE-SC0018623, DE-SC0020415, DE-SC0021311, DE-SC0021325,
DE-SC0021622, DE-SC0021629, DE-SC0021657, DE-SC0022012,
DE-SC0022270, DE-SC0022272, and DE-SC0023289, the Ida M. Green fellowship, NSF GRFP grant 2141064, ``la Caixa" Foundation fellowship LCF/BQ/AA20/11820045, Mauricio and Carlota Botton Foundation fellowship, Commonwealth Fusion Systems, and Eni. Disclaimer: This report was prepared as an account of work in part sponsored by an agency of the United States Government. Neither the United States Government nor any agency thereof, nor any of their employees, makes any warranty, express or implied, or assumes any legal liability or responsibility for the accuracy, completeness, or usefulness of any information, apparatus, product, or process disclosed, or represents that its use would not infringe privately owned rights. Reference herein to any specific commercial product, process, or service by trade name, trademark, manufacturer, or otherwise does not necessarily constitute or imply its endorsement, recommendation, or favoring by the United States Government or any agency thereof. The views and opinions of authors expressed herein do not necessarily state or reflect those of the United States Government or any agency thereof.


\section{References}
\printbibliography[heading=none]


\appendix
\newpage
\section{Magnet Design and Device Maintenance}

\subsection{Derivation of Equation \ref{eq:mag_ramp_limit}} \label{subsec:mag_ramp_deriv}
Beginning with Eq. 6 from \cite{kim2013numerical}:
\begin{equation}
    L\frac{dI_{\rm op}}{dt} + V_c\left(\frac{I_{\rm op}}{I_c}\right)^n = R_c(I_{\rm ps} - I_{\rm op})
\end{equation}
where $L$ is the coil inductance, $I_{\rm op}$ is the current driven through the superconductor, $I_{\rm ps}$ is the current input through the powersupply, $I_c$ is the critical current, $V_c$ is the voltage criterion for $I_c$, $n$ is the number of turns in the coil, and $R_c$ is the characteristic resistance of the radial pathway. For a magnet with many turns and operating with a reasonable margin away from the critical current, $(\frac{I_{\rm op}}{I_c})^n\approx 0$, so the second term on the left hand side can be neglected, giving:
\begin{align}
    L\frac{dI_{\rm op}}{dt} = R_c(I_{\rm ps} - I_{\rm op}) \label{eq:simplified_circuit}
\end{align}
From Ohm's law, the following inequalities must be satisfied to avoid heating the magnet:
\begin{align}
    R_c(I_{\rm ps} - I_{\rm op})^2\leq P_{\rm cool}\\
    I_{\rm ps} - I_{\rm op}\leq \sqrt{\frac{P_{\rm cool}}{R_c}} \label{ineq:ips_isc}
\end{align}
where $P_{\rm cool}$ is the available cooling power. Plugging inequality \ref{ineq:ips_isc} into equation \ref{eq:simplified_circuit} gives:
\begin{align}
    \frac{dI_{\rm op}}{dt} \leq \frac{\sqrt{R_cP_{\rm cool}}}{L}
\end{align}

\subsection{Finite element model of the Central Solenoid and Poloidal Field Coils.}
\label{appx:cspf}

\subsubsection{Geometry of the CS/PF model}

The physical dimensions and currents of the six poloidal field coils were chosen to produce a suitable magnetic equilibrium. The dimensions of the central solenoid are a trade off between the space available to the solenoid (3.2 m diameter) and the maximum achievable field within structural and electromagnetic limits, as explained in \ref{sec:magnets}. These design parameters resulted in  0.6 m wide by 7.76 m tall CS with 3296 turns.

\begin{figure*}
    \centering
    \includegraphics[width=0.8\textwidth]{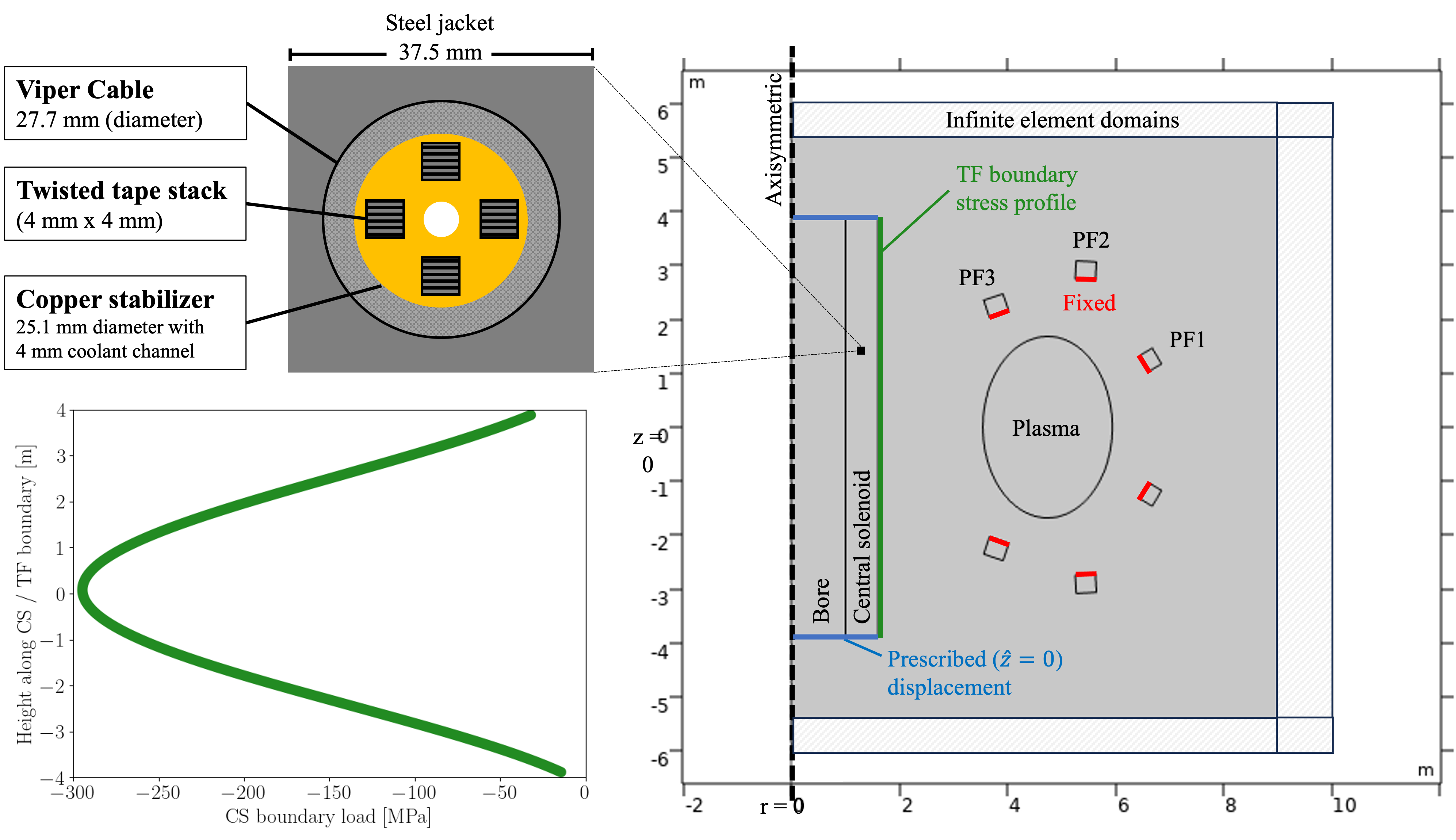}
    \caption{The dimensions and boundary conditions of the Central Solenoid (CS) and Poloidal Field (PF) Coils were set up to produce a full-account of Lorentz stresses and strain, including the effect of a 10 MA plasma and the boundary load on the outer edge of the CS resulting from the hoop stress in the toroidal field (TF) coils.}
    \label{fig:cspf}
\end{figure*}

Both sets of coils are modeled as COMSOL \textit{homogenized multi-turn coils}. The number of turns is determined by assuming a rectangular arrangement of turns, which have a square area of 1406 mm$^2$, accounting for the structural steel jacket supporting the VIPER cable. The VIPER cable is illustrated in Fig \ref{fig:cspf} and comprises four square twisted REBCO tape stacks (16 mm$^2$) embedded in a copper matrix. The copper enables current sharing between the stacks and thermal stability by exchanging heat with the central coolant duct\cite{hartwig2020viper}. The current carrying cross-section of a turn is the area of the VIPER cable, or 603 mm$^2$. The choice of current density per turn in the CS and PF coils is explained in section \ref{sec:magnets}. 

The COMSOL calculations also include the effect of a 10 MA plasma current, modeled as an elliptical conductor (1.2 x 1.6 m) with uniform current density.
\subsubsection{Materials and boundary conditions}

The coils are modeled as a mixed material with 8.5\% Cu and 91.5\% steel accounting for the fraction of copper in the VIPER cable, and assuming all other parts are made of steel. The bore of the CS is filled with structural steal as a simplest-possible solution to counter the hoop stresses from the TF and CS. A more comprehensive modeling of the CS should nevertheless consider the potential large AC losses incurred by this choice.

The boundary conditions of the model are represented on Fig. \ref{fig:cspf}. Most importantly, the model is axisymmetric about the center of the CS, and the outer boundaries of the air volume surrounding the CS and PF are infinite element domains that extrapolate magnetic field lines to infinity, avoiding a nonphysical closure of the field lines within the simulation volume.

The six PF coils are considered attached to the vacuum vessel on the inner side, and a prescribed ($\hat{z}=0$) displacement is imposed on the bottom and top of the CS and its supporting structure. While the toroidal field does not interact with poloidal coils, a compressive stress profile arises from the hoop force in the TF coils. The resulting radial force is modeled as a stress profile (see Fig. \ref{fig:cspf}) on the outer boundary of the CS.

\section{Balance of Plant}
\label{appx:thermo}
\subsection{Thermodynamic Analysis of Rankine Cycles}
A diagram of the Rankine cycle plant layout is presented in figure \ref{fig:brayton_cycle}. A full list of the thermodynamic states used in the final analysis is provided in table \ref{tab:brayton_thermo_states}. During analysis, all components were assumed to have an isentropic efficiency, $\eta_s = 0.95$. The electrical efficiency of the pumps is assumed to be $\eta_{pumps} = 0.75$. 
\begin{figure}[h]
    \centering
    \includegraphics[width=\columnwidth]{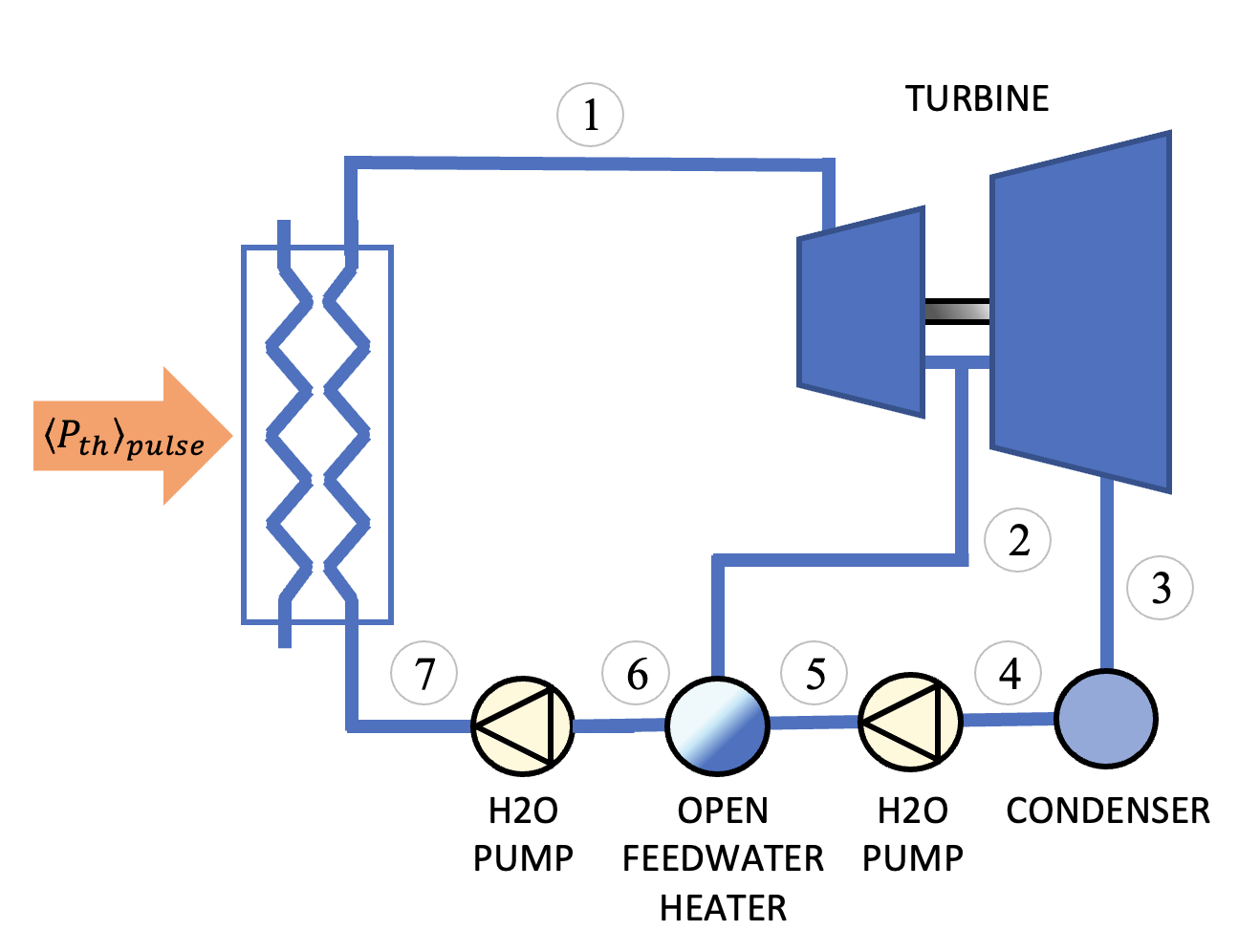}
    \caption{Diagram of Rankine cycle layout for sub and supercritical cycles. Heat from the secondary salt loop enters between states 7 and 1 as $\langle P_{th}\rangle_{pulse}$. A fraction of the working fluid is tapped at state 2, between high and low pressure turbine stages, for injection into the open feedwater heater. This reduces the total required pumping power. }
    \label{fig:rankine_cycle}
\end{figure}

\begin{table}
  \centering
  \begin{tabular}{lllll}
    \hline
    \multirow{2}{*}{State} & \multicolumn{2}{l}{Pressure [bar]} & \multicolumn{2}{l}{Temperature [\degree C]}\\
    \cline{2-5}
    & Sub & Super & Sub & Super\\
 \hline
 \hline
  1 & 150 & 300 & 560 & 560 \\ 
  2 & 23 & 52 & 292 & 297 \\ 
  3 & 0.6 & 0.6 & 86 & 86 \\ 
  4 & 0.6 & 0.6 & 86 & 86\\
  5 & 23 & 52& 86 & 86\\ 
  6 & 23 & 52& 220 & 266\\ 
  7 & 150 & 300& 223 & 274  \\ 
  \hline
    \end{tabular}
    \caption{Thermodynamic states for final subcritical and supercritical H$_\text{2}$O Rankine cycles. State labels correspond to those presented in figure \ref{fig:rankine_cycle}. Note that the temperature at state 5 is slightly higher than at state 4, but this has been obscured during rounding. }
    \label{tab:rankine_thermo_states}
\end{table}

\subsection{Thermodynamic Analysis of Brayton Cycle}
\begin{figure}[h]
    \centering
    \includegraphics[width=\columnwidth]{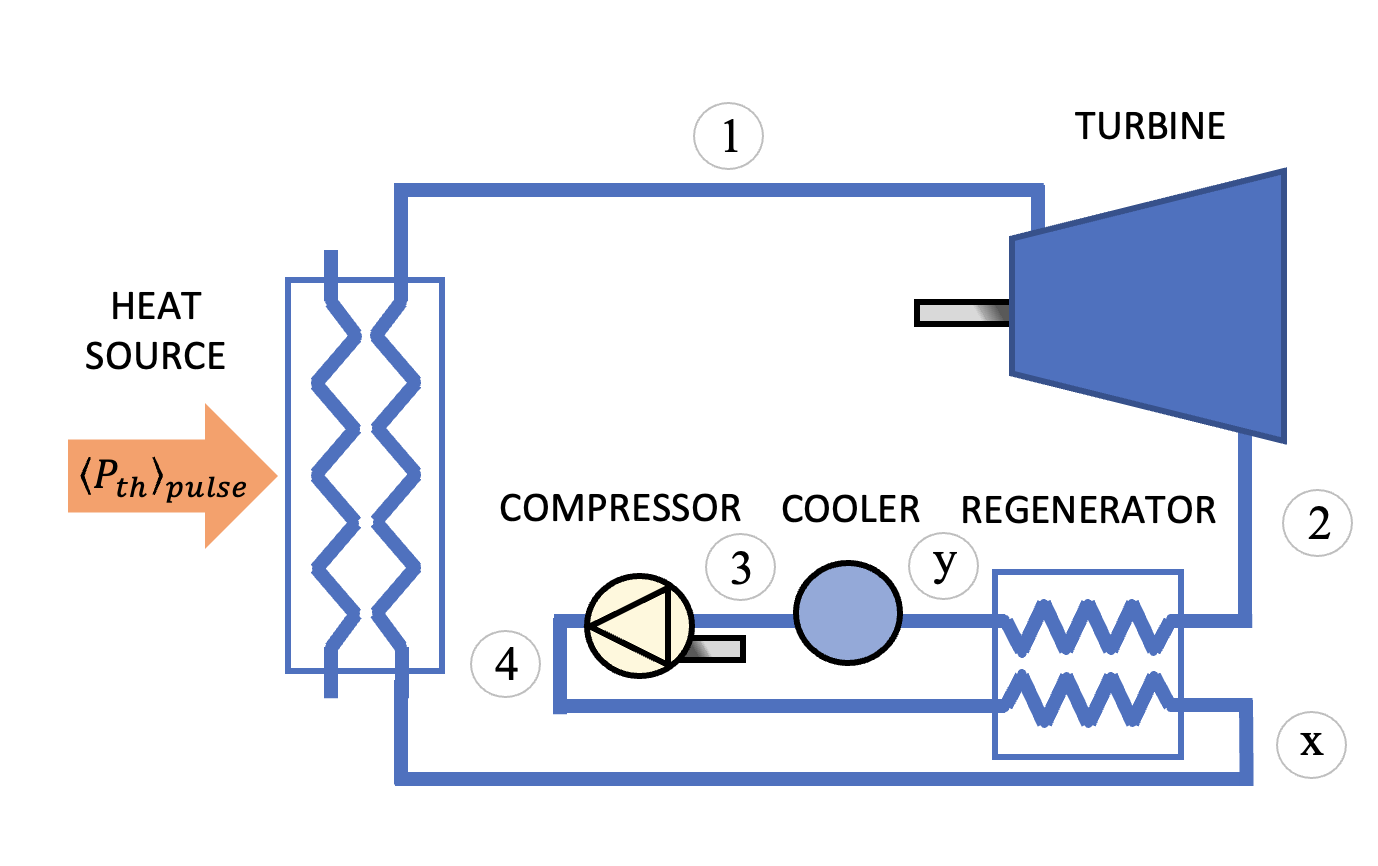}
    \caption{Diagram of Brayton cycle layout. Heat from the secondary salt loop enters between states 7 and x as $\langle P_{th}\rangle_{pulse}$. After exiting the turbine, Helium retains higher temperature than the compressor entrance, so it can be used for regeneration via a heat exchanger with outlet states x and y. For a given compression ratio, $p_4/p_3$, and maximum temperature, $T_1$, this reduces the amount of thermal power required to reach the same output power. Equivalently, this increases thermal efficiency of the cycle. Although not depicted here, the compressor can be mounted on the turbine shaft, so mechanical power can be directly tapped without intermediary losses due to electrical conversion. }
    \label{fig:brayton_cycle}
\end{figure}

A diagram of the Brayton cycle plant layout is presented in figure \ref{fig:brayton_cycle}. A full list of the thermodynamic states used in the final analysis is provided in table \ref{tab:brayton_thermo_states}.  During analysis, all components were assumed to have an isentropic efficiency, $\eta_s = 0.95$. The compressor is mounted directly to the turbine output shaft, and thus does not have an associated electrical efficiency. The optimized compression ratio, $p_4/p_3 = 2.1$. Finally, the regeneration efficacy is assumed to be $\eta_{regen} = 0.8$.

\begin{table}[h]
    \centering
    \resizebox{\columnwidth}{!}{
    \begin{tabular}{ lll }
 \hline
  \makecell{State}& \makecell{Pressure [bar]} & \makecell{Temperature [\degree C]} \\
 \hline
 \hline
  1 & 2.1 & 560  \\ 
  2 & 1.0 & 372  \\ 
  3 & 1.0 & 27  \\ 
  4 & 2.1 & 27  \\
  x & 2.1 & 325  \\ 
 \hline
    \end{tabular}
    }
    \caption{Thermodynamic states for final helium Brayton cycle. The working fluid is treated as an ideal gas. State labels correspond to those presented in figure \ref{fig:brayton_cycle}.}
    \label{tab:brayton_thermo_states}
\end{table}
%
%
{\onecolumn 
\section{Economic Analysis}

\subsection{Overnight Costing Methods}
\label{sec:overnight_costing}

\begin{table}[H]
\caption{Reactor plant}
\centering
\begin{tabular}{ m{3.5cm}  m{2.5cm} m{11cm} } 
    \hline 
  
  Component & Cost (US\$M) & Costing Method \\ 
  \hline \hline
  
  Toroidal Field Coils &  1,500  & Broken into tape, support structure, resistive lead, and power supply costs. A fabrication factor of 3x was applied to the support structure, and fabrication factor of 5x was applied to the REBCO tape and resistive leads. Inconel 718 used for structural supports. Necessary tape length and structural support materials were calculated by treating the magnets as a rectangle with rounded corners. \\
  
   \vspace{1ex} Poloidal Field Coils &  \vspace{1ex}180 &  \vspace{1ex}Broken into tape and support structure cost. A fabrication factor of 3x was applied to the support structure, and fabrication factor of 5x was applied to the REBCO tape. Nitronic 60 was used for structural supports. \\
  
   \vspace{1ex}Central Solenoid &  \vspace{1ex}140 & 
   \vspace{1ex}Broken into tape and support structure cost. A fabrication factor of 3x was applied to the support structure, and fabrication factor of 5x was applied to the REBCO tape. Nitronic 60 was used for structural supports. \\
  
   \vspace{1ex}Cryosystem &   \vspace{1ex}110 &  \vspace{1ex} Cryostat modeled as a 316 stainless steel cylinder containing the toroidal field coils. Its mass is scaled to ITER's, using the ratio of their surface areas\supercite{noauthor_ITERcryostat_nodate, noauthor_ITERmagnets_nodate}. For the cryocoolers, \$ 1M / kW was assumed for a total of \$90M \supercite{golfinopolous_notitle_2022}.\\
  
   \vspace{1ex}Remote Maintenance &  \vspace{1ex}55 &  \vspace{1ex}Value of contract for ITER's remote maintenance system ~\supercite{ITER_remote_maintenance}. The design of a such a system for a fusion power plant is very uncertain; this would be a valuable area for future research.\\
  
   \vspace{1ex}Plasma Heating &  \vspace{1ex} 370 &  \vspace{1ex} A cost of \$10M / MW was assumed for the ICRH system. \\
  
   \vspace{1ex}Divertor &  \vspace{1ex}150&  \vspace{1ex}Estimation based on previous devices.\\ 
  
   \vspace{1ex}Vacuum Vessel &  \vspace{1ex}23 & \vspace{1ex} Vacuum vessel modeled as two concentric rectangles of V-4Cr-4Ti, with rounded corners, and a fabrication factor of 3x. \\ 
  
   \vspace{1ex} Diagnostics & \vspace{1ex} 100 & \vspace{1ex} Extrapolated from existing radiation hardened diagnostics. \\
  
   \vspace{1ex} Blanket & \vspace{1ex} 380 & \vspace{1ex} Blanket geometry modeled as shells. Tank is made of Cr-V steel and filled with FLiBe. Thermal shielding and neutron shielding use Boron Carbide and Titanium Carbide respectively. Solid layers have a fabrication factor of 3x.\\
  
  \vspace{1ex} Tritium &   \vspace{1ex} 57&  \vspace{1ex}  \vspace{1ex} Assumed \$30M for tritium handling system \cite{woodruff_conceptual_2017}. For start up inventory of 900g, a cost of \$30,000/g was taken\supercite{UKAEA_Tritium_18}. \\
  \hline 

  

\end{tabular}
\label{tab:componentCosts}
\end{table}

\begin{table}[H]
\caption{Direct costs}
\centering
\begin{tabular}{  m{4cm}  m{3cm} m{10cm}  } 
  \hline \
  Item & Cost (US\$M) & Costing Method \\ 
  \hline \hline
  Reactor equipment & & Sum from table \ref{tab:componentCosts}. \newline \\
  Turbine plant &  200 & Inflation-corrected estimate for power generation using a first of a kind, 350MWt, Rankine cycle on a high temperature gas cooled reactor \supercite{INL_HTGR_costing}. Assumed interpulse storage costs are \$25/kWh \supercite{forsberg_saltcost_2022}, resulting in a negligible additional cost. Estimate is conservative w.r.t a small fission reactor\supercite{EIA_Capital_Costs_2013}.\\
  Electric plant & 35 & \vspace{2ex} CBS 24 in \cite{woodruff_conceptual_2017}. \newline \\
  Misc plant & 17 & \vspace{2ex} CBS 25 in \cite{woodruff_conceptual_2017}. \newline \\
  Heat rejection & 11 & \vspace{2ex} CBS 26 in \cite{woodruff_conceptual_2017}. \newline \\
  Land & 15 & \vspace{2ex} CBS 20 in \cite{woodruff_conceptual_2017}. \newline \\
  Structures & 110 & \vspace{2ex} CBS 21 in \cite{woodruff_conceptual_2017}. \newline \\
  \hline
\end{tabular}
\label{tab:Direct costs}
\end{table}

\begin{table}[h]
\caption{Indirect costs}
\centering
\begin{tabular}{  m{5cm}  m{3cm} m{9cm}  } 
  \hline \
  Item & Cost (US\$M) & Costing Method \\ 
  \hline \hline
  Construction services & 75 & \vspace{2ex} CBS 91 in \cite{woodruff_conceptual_2017}. \newline \\
  Home office engineering & 18 & \vspace{2ex} CBS 92 in \cite{woodruff_conceptual_2017}. \newline \\
  Field office engineering & 38 & \vspace{2ex} CBS 93 in \cite{woodruff_conceptual_2017}. \newline \\
  Owner's costs & 18 & \vspace{2ex} CBS 94 in \cite{woodruff_conceptual_2017}. \newline \\
  \hline
\end{tabular}
\label{tab:Indirect Costs}
\end{table}

\subsection{Materials Costs}
\label{sec:materials_costs}
\begin{table}[H]
\caption{Material costs}
\centering
\begin{tabular}{  m{4cm}  m{4.25cm} || m{4cm}  m{4.25cm}  } 
  \hline \
  Component & Cost & Component & Cost \\
  \hline \hline
  
  REBCO & \$ 40/kAm [assumed]& Titanium Carbide&  \$85/kg \supercite{BC_price} \\[1ex]
  TF Power Supply & \$0.5M/supply \supercite{mouratidis_manta_2023} & Stainless steel 316& \$6.5 /kg\supercite{noauthor_metal_nodate}\\[1ex]
  TF Resistive Lead& \$2 M/lead [assumed]& Inconel 718& \$65/kg \supercite{noauthor_inconel_nodate}\\[1ex]
  Deuterium& \$21/g \supercite{Deuterium_Price_2} & Nitronic 60& \$25/kg \supercite{noauthor_nitronic_nodate}\footnotemark[3]\\[1ex]
  Tritium& \$30k/g \supercite{science_tritium_price}& Cr-V Steel& \$43/kg \supercite{arainejad_notitle_2023}\\[1ex]
 FLiBe& \$169/kg \supercite{ARC_2015} &  &\\[1ex]
 Boron Carbide & \$142/kg  \supercite{BC_price} &  & \\
 \hline
\end{tabular}
\label{tab:materials costs}
\end{table}

\footnotetext[3]{Price assumed 50\% higher in US than India}


\begin{subsection}{Operational Costing: LCOE Terms}
    \label{sec:A_Operational_Costing}
    \begin{table}[H]
    \caption{Operational costs}
    \centering
    \begin{tabular}{  m{4.5cm}  m{12.5cm}  } 
      \hline \
      
      Cost & Quantity and Calculation Method \\ 
      \hline \hline
            Financing & A 3\% interest rate, 3\% inflation, 0\% down-payment, and that the entire loan for the capital cost (including plant, facilities, and decommissioning) is payed off by the end of the project was assumed.\\
            

             \vspace{2ex}Taxes \& Tax Credits & \vspace{2ex}Taxes paid or tax credits received vary too much by location to be usefully included. Note that MANTA would likely be eligible for a nontrivial value of production tax credits such as under the Inflation Reduction Act\cite{IRA_2022}, at time of publication.\\
            
             \vspace{2ex}Personnel Costs &  \vspace{2ex}15\$M/yr = ($\sim$1 person/MWe)$\times$\$150,000/(employee-year) \supercite{IAEA_SMR_Staffing}\\
            
             \vspace{2ex}Yearly Capital Repairs & 
             \vspace{2ex} 0.5\% of capital cost  + 1 vacuum vessel every two years (including fabrication and removal/installation) due to radiation damage.\\
            
             \vspace{2ex}Magnet Replacement & 
             \vspace{2ex} It was assumed that only the REBCO portion of the magnet requires replacement due to radiation damage, and the replacement times (see table \ref{tab:maglifetimes}) are grouped to the nearest half-year to decrease the number of zero-production maintenance months. A 2 month maintenance downtime was assumed, and a REBCO replacement cost ranging from \$853M for the TF magnet to \$20.9M-\$36.9M for the PFs. \\
            
             \vspace{2ex}Fueling Costs & 
              \vspace{2ex}After a 900g startup inventory, fueling is defined as a negative cost due to selling tritium at an assumed current price of \$30k/g\supercite{science_tritium_price}.\\

              \vspace{2ex}Electricity Cost & \vspace{2ex}\$47.6/MWhr\supercite{EIA_Wholesale_electricy} is used as the US east coast wholesale marginal price for selling electricity. \\ 

             \vspace{2ex} Power Production &  
              \vspace{2ex}An effective duty cycle of 90\% was calculated due to recharging the thermal reservoir to maintain constant electrical producing during CS swings, a further 88\% for plant availability due to maintenance downtime (2 month per magnet replacement), 36\% for turbine efficiency, and 95\% for each of the four Heat Exchangers (not always simultaneously in use, see Fig. \ref{fig:molten_salt_storage}). Disruptivity and uncontrolled shutdowns are not included in the analysis (other works have investigated these effects on plant economic viability\supercite{Maris_2023}). Total thermal power is calculated in Sec. \ref{sec:balance_of_plant}. \\
            
             \vspace{2ex}Power Consumption &  
             \vspace{2ex}Electrical consumption consists of auxiliary heating (40 MW at a wall-plug efficiency of 70\%), cryostat (1.5 MW\supercite{kittel_2007}), turbine and FLiBe pump (4.88 MW), and diagnostic systems (0.1 MW).\\
            \hline
        \end{tabular}
        
    \label{tab:Operational_Costs}
    \end{table}
\end{subsection}

}
\end{document}